\documentstyle[12pt]{article}

\input epsf.sty

\newcommand{\inclfig}[2]{\mbox{\epsfxsize=#1cm \epsfbox{#2.ps}}}
\newcommand{\insertfig}[2]{\mbox{\epsfxsize=#1cm \epsfbox{#2.eps}}}

\newcommand{\cO}{{\cal O}}
\newcommand{\cP}{{\cal P}}
\newcommand{\cM}{{\cal M}}
\newcommand{\cQ}{{\cal Q}}

\begin{document}

\begin{titlepage}

\centerline{\large \bf Leading twist asymmetries in deeply virtual
                       Compton scattering.}

\vspace{15mm}

\centerline{\bf A.V. Belitsky$^a$, D. M\"uller$^{a,b}$,
                L. Niedermeier$^b$, A. Sch\"afer$^b$}

\vspace{10mm}

\centerline{\it $^a$C.N.\ Yang Institute for Theoretical Physics}
\centerline{\it State University of New York at Stony Brook}
\centerline{\it NY 11794-3840, Stony Brook, USA}

\vspace{5mm}

\centerline{\it $^b$Institut f\"ur Theoretische Physik,
                Universit\"at Regensburg}
\centerline{\it D-93040 Regensburg, Germany}

\vspace{15mm}

\centerline{\bf Abstract}

\vspace{0.5cm}

We calculate spin, charge, and azimuthal asymmetries in deeply virtual
Compton scattering at leading twist-two level. The measurement of these
asymmetries gives access to the imaginary and real part of all deeply
virtual Compton scattering amplitudes. We note that a consistent description
of this process requires taking into account twist-three contributions and
we give then a model dependent estimate of these asymmetries.

\vspace{5cm}


\vspace{0.5cm}


\end{titlepage}

\section{Introduction.}

The scattering of electroweak probes off hadrons serves as a clean tool,
free of complications from hadron-hadron reactions on both the theoretical
and experimental sides, for the extraction of reliable information on the
substructure of strongly interacting particles. Using the photon as a
probe, (the absorptive part of) the forward virtual Compton (VC) process
$\gamma^*(q) N (P) \to \gamma^*(q) N(P)$ allows to study the strong
interaction dynamics and at the same time it has a simple QCD description
in the hard regime, --- when $- q^2 \gg m^2_{\rm hadr}$. In more general
settings it is instructive to address the non-forward scattering
$\gamma^{(*)}(q_1) N (P_1) \to \gamma^{(*)}(q_2) N(P_2)$. The systematic
approach to its calculation is established only for the deep inelastic
domain, where the QCD factorization theorems separate short and long
distance phenomena into a perturbative parton subprocess and soft
functions which encode information about the strongly coupled regime. The
latter, known as the skewed parton distributions (SPDs), being studied for
some time \cite{MueRobGeyDitHor94}, have attracted increased attention in
light of the conceivable opportunity to learn more about the spin
structure of the nucleon \cite{Ji97}. They are also of interest in their
own right being hybrids of parton densities/distribution amplitudes and
form factors. They share properties of the former in different regions of
phase space \cite{Rad96} which have been studied perturbatively to a great
extent at one and two-loop orders \cite{BelMul98}. Deeply virtual Compton
scattering (DVCS), with $q_2^2 = 0$, proves to be an experimentally
accessible reaction \cite{Sau99}. In electroproduction processes of a real
photon there is a strong contamination of DVCS by the Bethe-Heitler (BH)
process. In view of the extreme interest to extract, or at least to
constrain, SPDs it is timely to address the question of the best
observables that allow to get rid of unwanted background. Fortunately,
the interference of the two processes provides a rich source of
information. It was suggested that diverse asymmetries
\cite{KroSchGui96,Ji97,DieGouPirRal97,FraFreStr97,FraFreStr98,FreStr99}
can be used to disentangle the real and imaginary parts of DVCS and thus
give access to SPDs. In the present contribution we consider a number of
spin, azimuthal and charge asymmetries which share these properties and
give predictions for the kinematics of the HERA and HERMES experiment.
Spin asymmetries make it possible to extract the imaginary part of the
DVCS amplitude and thus, due to the reality of SPDs, which holds owing
to the spatial and time reversal invariance of strong interactions, give
directly a measurement of the shape (at leading order in $\alpha_s$ in
complete analogy to DIS) of SPDs on the diagonal $t = \xi$.

The paper is organized as follows. In section \ref{Sec-CrosSec} we
calculate the squared amplitude for DVCS, BH, and the interference term
to leading twist-two accuracy. In section \ref{Sec-ExtofAmp} we present
simplified formulae for different cross sections and give a qualitative
discussion of the feasibility to measure the DVCS leading twist-two
amplitudes in different kinematical regions. After introducing models
for SPDs, we give an estimate for different asymmetries for HERA and
HERMES kinematics in \ref{Sec-NumEst}. Finally, in section \ref{Sec-Con}
we summarize.

\section{Cross section.}
\label{Sec-CrosSec}

To start let us discuss different contribution to the differential
cross section of the electroproduction process $e(k,\lambda) N(P_1,S_1)
\to e(k^\prime,\lambda) N(P_2,S_2)\gamma(q_2,\Lambda)$, given by the
standard formula
\begin{eqnarray}
\label{cross-section-def}
d \sigma = \frac{1}{4 k.P_1} |{\cal T}|^2(\lambda,S_1)
(2\pi)^4 \delta^4( k + P_1 - k^\prime - P_2 - q_2 )
\frac{d^3 \mbox{\boldmath$k$}^\prime}{2 \omega^\prime (2\pi)^3}
\frac{d^3 \mbox{\boldmath$P$}_2}{2 E_2 (2 \pi)^3}
\frac{d^3 \mbox{\boldmath$q$}_2}{2 \nu (2 \pi)^3} .
\end{eqnarray}
\begin{figure}[t]
\begin{center}
\vspace{-0.3cm}
\hspace{1cm}
\mbox{
\begin{picture}(0,300)(300,0)
\put(100,140){\inclfig{20}{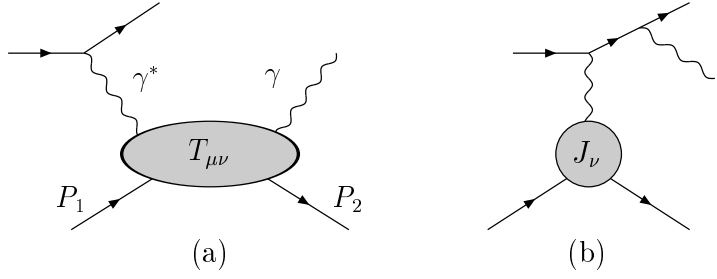}}
\end{picture}
}
\end{center}
\vspace{-7.5cm}
\caption{\label{electroprod} The virtual Compton scattering amplitude
and the Bethe-Heitler process.}
\end{figure}

The scattering amplitude squared $|{\cal T}|^2$ in the cross section,
contains beside the VCS [Fig.\ \ref{electroprod}(a)] and BH [Fig.\
\ref{electroprod}(b) and crossed contribution] parts also the interference
term:
\begin{eqnarray}
|{\cal T}|^2(\lambda,S_1)
= \sum_{\lambda^\prime, S_2, \Lambda}
\left\{
|{\cal T}_{VCS}|^2 + |{\cal T}_{BH}|^2
+ {\cal T}_{VCS} {\cal T}_{BH}^\ast
+ {\cal T}_{VCS}^\ast {\cal T}_{BH}
\right\}.
\end{eqnarray}
The BH-amplitude is purely real and is given as a contraction of the
leptonic tensor, at leading order in the fine structure constant $\alpha$,
\begin{equation}
L_{\mu\nu}
= \bar u (k^\prime, \lambda^\prime)
\left[
\gamma_\mu (\slash\!\!\! k - \slash\!\!\!\!\Delta)^{-1} \gamma_\nu
+ \gamma_\nu (\slash\!\!\! k^\prime + \slash\!\!\!\!\Delta)^{-1} \gamma_\mu
\right]
u(k, \lambda) ,
\end{equation}
with the hadronic current
\begin{eqnarray}
\label{def-HadCur}
J_\nu = \bar U (P_2,S_2)
\left\{ F_1(\Delta^2) \gamma_\nu
+ i F_2(\Delta^2) \sigma_{\nu\tau} \frac{\Delta^\tau}{2M}
\right\} U (P_1,S_1) ,
\quad \mbox{where}
\quad \Delta = P_2 - P_1 = q_1 - q_2,
\end{eqnarray}
parametrized in terms of Dirac, $F_1$, and Pauli, $F_2$, form factors
normalized according to $F_1^p (0) = 1$, $F_2^p (0) \equiv \kappa_p =
1.79$, and $F_1^n (0) = 0$, $F_2^n (0) \equiv \kappa_n = - 1.91$, for
proton and neutron, respectively. Thus, the BH amplitude is of the form
\begin{eqnarray}
\label{BH-amplitude}
{\cal T}_{BH} = - \frac{e^3}{\Delta^2} \epsilon^\ast_\mu L^{\mu\nu} J_\nu .
\end{eqnarray}
The form factors are known fairly well from experimental measurements
and can be pa\-ra\-met\-rized by dipole formulae in the small $\Delta^2$
region
\begin{equation}
\label{dipoleFF}
G_E^p (\Delta^2)
= (1 + \kappa_p)^{-1} G_M^p (\Delta^2)
= \kappa_n^{-1} G_M^n (\Delta^2)
= \left( 1 - \frac{\Delta^2}{m_V^2} \right)^{- 2} ,
\qquad
G_E^n (\Delta^2) = 0,
\end{equation}
where we have introduced the electric, $G_E^i (\Delta^2) = F_1^i (\Delta^2)
+ \frac{\Delta^2}{4 M^2} F_2^i (\Delta^2)$, and magnetic, $G_M^i (\Delta^2)
= F_1^i (\Delta^2) + F_2^i (\Delta^2)$, form factor characterized by
cutoff mass $m_V = 0.84\, {\rm GeV}$, see e.g.\ \cite{Oku90}. The hadronic
tensor $T_{VCS}$ is:
\begin{eqnarray}
\label{VCS-amplitude}
{\cal T}_{VCS}
= \mp \frac{e^3}{q_1^2} \epsilon^\ast_\mu T^{\mu\nu}
\bar u (k^\prime) \gamma_\nu u(k),
\quad\mbox{where}\quad
\left\{ {- \mbox{\ for\ } e^- \atop + \mbox{\ for\ } e^+ } \right. .
\end{eqnarray}
It is defined by the time ordered product of two electromagnetic currents
\begin{eqnarray}
T_{\mu\nu} (q, P_1, P_2) =
i \int dx e^{i x.q}
\langle P_2, S_2 | T j_\mu (x/2) j_\nu (-x/2) | P_1, S_1 \rangle,
\end{eqnarray}
where $q = (q_1 + q_2)/2$ (and the index $\mu$ refers to the outgoing
real photon). It contains for a spin-1/2 target twelve\footnote{$12
= \frac{1}{2} \times 3 \mbox{\ (virtual photon)} \times 2 \mbox{\ (final
photon)} \times 2 \mbox{\ (initial nucleon)} \times 2 \mbox{\ (final
nucleon)}$. The reduction factor $1/2$ is a result of parity invariance.}
independent kinematical structures \cite{KroSchGui96}. In this paper we
restrict ourselves to the twist-2 part of $T_{\mu\nu}$ that does not contain
transversal photon spin flip contributions. Such contributions arise in the
next-to-leading order of perturbation theory due to the gluon transversity
\cite{DieGouPirRal97,HooJi98} and are seperately considered in
\cite{BelMue00}. From the structure of the OPE we immediately learn that
these contributions are contained in the following form factor
decomposition\footnote{We adopt throughout the conventions of
\cite{ItzZub}, e.g.\ $\epsilon^{0123} = 1$}:
\begin{eqnarray}
\label{decom-T}
T_{\mu\nu} (q,P,\Delta)
&=&
- \tilde{g}_{\mu\nu} \frac{q_\sigma V_1^\sigma}{P.q}
- i \tilde{\epsilon}_{\mu \nu q \sigma} \frac{A_1^\sigma}{P.q}
+ \cdots ,
\end{eqnarray}
where $P = P_1 + P_2$ and the gauge invariant tensors $\tilde t_{\mu\nu}
= \cP_{\mu\rho} t_{\rho\sigma} \cP_{\sigma\nu}$ are constructed by means
of the projection tensor $\cP_{\mu\nu} \equiv g_{\mu\nu} - q_{1\,\mu}
q_{2\,\nu}/q_1.q_2$. The ellipsis indicate twist-three and higher
contributions. The vectors $V_{i\mu}$ and axial-vectors $A_{i\nu}$ can
be expressed by a form factor decomposition
\begin{eqnarray}
\label{dec-FF-V}
V_{1\mu}
= \bar U (P_2, S_2)
\left( {\cal H}_1 \gamma_\mu
+ {\cal E}_1 \frac{i\sigma_{\mu\nu} \Delta^\nu}{2M}
\right) U (P_1, S_1) + \cdots , \\
\label{dec-FF-A}
A_{1\mu}
= \bar U (P_2,S_2)
\left( \widetilde{\cal H}_1 \gamma_\mu\gamma_5
+ \widetilde{\cal E}_1 \frac{\Delta_\mu \gamma_5}{2M}
\right) U (P_1,S_1) + \cdots ,
\end{eqnarray}
where higher twist contributions are neglected. These form factors depend
on the following variables
\begin{eqnarray}
\xi = \frac{Q^2}{P.q},
\qquad
Q^2 = -q^2 = - \frac{1}{4}(q_1 + q_2)^2,
\qquad
\Delta^2.
\nonumber
\end{eqnarray}
Note that in general (for off-shell final photons) a second scaling variable
$\eta = \frac{\Delta.q}{P.q}$ appears, which is related however to $\xi$,
i.e.\ $\eta = -\xi \left(1 - \frac{\Delta^2}{4 Q^2}\right) \approx - \xi$.
The amplitudes are given as convolution in $t$, $\otimes \equiv \int dt$,
of perturbatively calculable hard scattering parts with SPDs:
\begin{eqnarray}
\left\{{{\cal H}_{1} \atop {\cal E}_1} \right\} (\xi, Q^2, \Delta^2)
\!\!\!&=&\!\!\!
T_1 (\xi, Q^2, \mu^2, t)
\otimes \left\{{H \atop E}\right\} (t, \xi, \Delta^2, \mu^2) , \\
\left\{
{\widetilde{{\cal H}}_{1} \atop {\widetilde{\cal E}}_1}
\right\}(\xi, Q^2, \Delta^2)
\!\!\!&=&\!\!\!
{\widetilde T}_1(\xi, Q^2, \mu^2,t) \otimes
\left\{{{\widetilde H}\atop {\widetilde E}}\right\}
(t, \xi, \Delta^2, \mu^2) ,
\end{eqnarray}
with summation over the different parton species implied and $\mu^2$ being
the factorization scale. The hard scattering amplitudes are available in
next-to-leading order (NLO) approximation and they read in LO for a quark
of charge $Q_i$
\begin{equation}
\xi\, {T^{i(0)}} \left( \xi, t \right)
= \frac{Q^2_i}{1 - t/\xi - i \epsilon}
\mp (t \to -t) ,
\end{equation}
with $-$ ($+$) for parity even (odd) cases.

\begin{figure}[t]
\begin{center}
\vspace{-0.3cm}
\hspace{1cm}
\mbox{
\begin{picture}(0,300)(300,0)
\put(110,90){\inclfig{27}{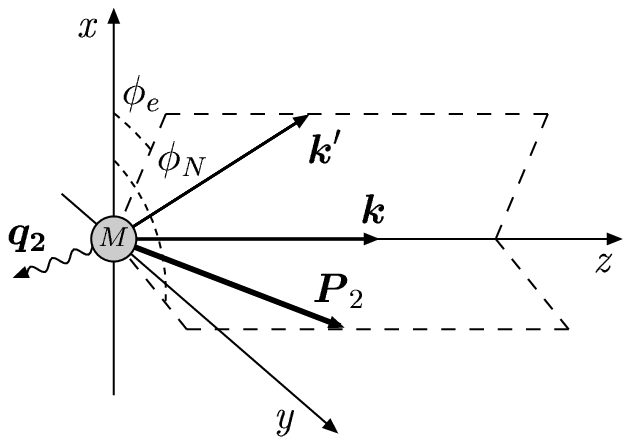}}
\end{picture}
}
\end{center}
\vspace{-6cm}
\caption{\label{Kinematics} The kinematics of the reaction
$e (\mbox{\boldmath$k$}) N (M) \to e(\mbox{\boldmath$k$}')
N (\mbox{\boldmath$P$}_2) \gamma (\mbox{\boldmath$q$}_2)$ in the
rest frame of the target.}
\end{figure}

In the consequent presentation we give our results in the laboratory frame
(see Fig.\ \ref{Kinematics}) in which we use the kinematical variables
\begin{eqnarray}
\label{Lframe}
&& k = (E, 0, 0, E), \quad
k^\prime = (E^\prime, E	^\prime \cos\phi_e \sin\theta_e,
E^\prime \sin\phi_e \sin\theta_e,E^\prime \cos\theta_e), \\
&& P_1 = (M, 0, 0, 0), \
P_2 = (E_2, |\mbox{\boldmath$P$}_2| \cos\phi_N \sin\theta_N,
|\mbox{\boldmath$P$}_2|\sin\phi_N \sin\theta_N,
|\mbox{\boldmath$P$}_2| \cos\theta_N).
\nonumber
\end{eqnarray}
Furthermore, we introduce the azimuthal angle $\phi_r = \phi_N - \phi_e$
between the lepton and hadron scattering planes as well as $\phi_s =
\phi_N + \phi_e$. The spin vector of the spin-1/2 target for longitudinal
and transverse polarization is given by
\begin{eqnarray}
S=(0, 0, 0, \Lambda)
\quad\mbox{with}\quad \Lambda = \pm 1,
\qquad
S=(0, \cos{\mit\Phi}, \sin{\mit\Phi}, 0),
\end{eqnarray}
respectively.

From the experimental point of view one works with the variables
${\cal Q}^2\equiv -q_1^2$ and $x \equiv - q_1^2/(2 P_1.q_1)$,
which are related to our variables by
\begin{eqnarray}
Q^2 = - \frac{1}{2} q_1^2 \left( 1 - \frac{\Delta^2}{2q_1^2} \right)
\approx \frac{1}{2} {\cal Q}^2,
\quad
\xi = \frac{x \left( 1 - \frac{\Delta^2}{2 q_1^2} \right)}
{2 - x \left( 1 + \frac{\Delta^2}{q_1^2} \right)}
\approx \frac{x}{2 - x} .
\end{eqnarray}
After performing the phase space integration we obtain for the
differential cross section (\ref{cross-section-def})
\begin{eqnarray}
\label{WQ}
\frac{d\sigma}{dx d{\cal Q}^2 d|\Delta^2| d\phi_r}
=
\frac{y}{{\cal Q}^2} \frac{d\sigma}{dx dy d|\Delta^2| d\phi_r}
=
\frac{\alpha^3  x y^2 } { 8 \, \pi \,  {\cal Q}^4}
\left( 1 + \frac{4 M^2 x^2}{{\cal Q}^2} \right)^{-1/2}
\left| \frac{\cal T }{e^3} \right|^2 ,
\end{eqnarray}
where we introduced as well the conventional variable $y = P_1.q_1/P_1.k$
[$y = 1 - E'/E$ in the frame (\ref{Lframe})].

Here we present simple analytical expressions for the amplitudes entering
the cross section for positron beam. Changing to electrons will generate a
minus sign for the interference terms. To deduce them we present at first
the amplitudes squared in terms of the form factors. The result for the
DVCS amplitude reads in leading twist-2 approximation (we set $|e| = 1$)
\begin{eqnarray}
|{\cal T}_{\rm DVCS}|^2 &=&
8 \frac{2 - 2y + y^2}{y^2} \frac{\xi^2}{{\cal Q}^6}
\left(q.V_1\ q.V_1^\dagger + q.A_1\ q.A_1^\dagger \right)
\nonumber \\
&&+ 8 \frac{\lambda (2-y)}{y} \frac{\xi^2}{{\cal Q}^6}
\left(q.A_1\ q.V_1^\dagger + q.V_1\ q.A_1^\dagger \right),
\end{eqnarray}
where $\lambda = 1$ means that the spin of the lepton is parallel to
the beam direction. It is obvious that the leading contribution scales as
$1/{\cal Q}^2$.

The exact squared amplitude for BH reads in terms of the
electromagnetic currents:
\begin{eqnarray}
|{\cal T}_{\rm BH}|^2 &=&
\frac{8}{\Delta^2} \frac{q.J\ k.J^\dagger + k.J\ q.J^\dagger
- q.J\ q.J^\dagger - 2 k.J\ k.J^\dagger }
{(2 k.\Delta -\Delta^2)({\cal Q}^2 + 2 k.\Delta)}
\nonumber\\
& &-\frac{4}{\Delta^4}
\left(1 + \frac{{\cal Q}^4 +  \Delta^4}
{2(2 k.\Delta -\Delta^2)({\cal Q}^2 + 2 k.\Delta)} \right)
J.J^\dagger
\\
& & +\frac{4 i\lambda
\left[\left({\cal Q}^2 + 4 k.\Delta -\Delta^2 \right)
\epsilon_{q \Delta J J^\dagger}
+ \Delta^2 \epsilon_{k (2 q + \Delta) J J^\dagger} \right]}
{\Delta^4 (2 k.\Delta -\Delta^2)({\cal Q}^2 + 2 k.\Delta)},
\nonumber
\end{eqnarray}
with the latter being defined in Eq.\ (\ref{def-HadCur}). Instead of the
exact expression we may use an approximated one. To be consistent we have
to expand the squared BH-amplitude, which starts with $1/\Delta^2$, and
the interference term up to the same order as the squared DVCS-amplitude,
namely to $1/{\cal Q}^2$ accuracy. However, one has to take into account
that the lepton propagator in the $u$-channel of the BH amplitude gives a
contribution which behaves as $(k - q_2)^2 = - \frac{(1-y)}{y} {\cal Q}^2
\left( 1 + \cO (1/{\cal Q}) \right)$. Therefore, the Taylor expansion is not
legitimate for large $y$ and sets, otherwise, an upper limit for $y$, namely
$1 - y \gg M^2/{\cal Q}^2$. To avoid this problem, we have to
expand the propagator,
\begin{eqnarray}
\label{app-prop}
(k - q_2)^2 &=&
- \frac{\cQ^2}{y}  \Bigg\{1 - y +
2\sqrt{\frac{-\Delta^2}{\cQ^2}}
\sqrt{1 -\frac{\Delta^2_{\rm min}}{\Delta^2}}
\sqrt{1 - y}\sqrt{1 - x}\cos(\phi_r)-
\frac{\Delta^2}{\cQ^2}\Bigg[ \frac{1-y}{2}\\
&&+
(1 - x)\left(1 - 2\frac{\Delta^2_{\rm min}}{\Delta^2}\right) -
(1 - y) \left(x + 2(1 - x)\frac{\Delta^2_{\rm min}}{\Delta^2}\right)
\cos(2 \phi_r) \Bigg] + O\left(\frac{1}{\cQ^3} \right)\Bigg\},
\nonumber
\end{eqnarray}
and the remaining parts separately in Taylor series which results in a
Pade-like approximation for the squared BH term. Here $\Delta_{\rm min}^2$
is the minimal value of $\Delta^2$ which is defined by the kinematical
restriction $\Delta_{\rm min}^2 = - M^2 x^2/ (1 - x + x M^2/\cQ^2)$.
Taking only the first terms in the expansion, thus, neglecting all
contributions formally suppressed by $1/{\cal Q}$, we get
\begin{eqnarray}
|{\cal T}_{\rm BH}|^2 &=& - \frac{2 ( 2 - 2 y + y^2 )}{(1 - y)}
\left(\frac{J.J^\dagger}{\Delta^4}
+ 4 \frac{q.J\ q.J^\dagger}{\Delta^2 {\cal Q}^4} \right)
- 4 \frac{\lambda ( 2 - y ) y^2}{( 1 - y )}
\frac{i \epsilon_{k \Delta J J^\dagger}}{\Delta^4 {\cal Q}^2},
\end{eqnarray}
which, however, is valid for $y \ll 1$.

For interference terms we adhere to the same approximation and give here
only the leading contributions in $1/{\cal Q}$, i.e.\
\begin{eqnarray}
{\cal T}_{\rm BH} {\cal T}_{\rm DVCS}^\ast
&=& - 4 \frac{2 - 2y + y^2}{1 - y} \frac{\xi}{\Delta^2 {\cal Q}^4}
\left[ \left( k^\sigma -\frac{1}{y} q^\sigma \right)
\left( J_\sigma + 2 \Delta_\sigma \frac{q.J}{{\cal Q}^2} \right)
q.V_1^\dagger
-2 i \epsilon_{k q \Delta J} \frac{q.A_1^\dagger}{{\cal Q}^2} \right]
\nonumber\\
&&- 4 \frac{\lambda(2 - y) y}{1 - y} \frac{\xi}{\Delta^2 {\cal Q}^4}
\left[ \left( k^\sigma -\frac{1}{y} q^\sigma \right)
\left( J_\sigma + 2 \Delta_\sigma \frac{q.J}{{\cal Q}^2} \right)
q.A_1^\dagger
- 2 i \epsilon_{k q \Delta J} \frac{q.V_1^\dagger}{{\cal Q}^2} \right].
\end{eqnarray}
Note that for a consistent expansion up to order $1/{\cal Q}^2$ also
twist-three contributions having both kinematical and dynamical origins
must be considered (see Ref.\ \cite{AniTerPir00} for the case of a scalar
target). In general these contributions are poorly understood and have to
be studied in more detail.

As a next step for the calculation of the cross section we have to sum over
the spin of the outgoing proton and write the final answer in terms of SPDs
and electromagnetic form factors. Here we give the results for polarized
spin-1/2 target, where $\Lambda = 1$ means polarization along the lepton
beam, averaged over the azimuthal angle $\phi_s$. The DVCS amplitude
squared $|{\cal T}_{\rm DVCS}|^2_{\rm} = |{\cal T}_{\rm DVCS}|^2_{\rm unp}
+ |{\cal T}_{\rm DVCS}|^2_{\rm pol}$, with ${\rm pol} = \{{\rm LP, TP}\}$,
consists of the elements
\begin{eqnarray}
\label{def-DVCS-unp}
|{\cal T}_{\rm DVCS}|^2_{\rm unp}
\!\!\!&=&\!\!\! \frac{2 ( 2 - 2 y + y^2 )}{y^2 (2 - x)^2 {\cal Q}^2}
\Bigg[ 4 (1 - x)
\left(
{\cal H}_1 {\cal H}_1^\ast
+ \widetilde{\cal H}_1 \widetilde {\cal H}_1^\ast
\right)
- x^2
\bigg(
{\cal H}_1 {\cal E}_1^\ast + {\cal E}_1 {\cal H}_1^\ast
+ \widetilde{{\cal H}}_1 \widetilde{{\cal E}}_1^\ast
+ \widetilde{{\cal E}}_1 \widetilde{{\cal H}}_1^\ast
\bigg)
\nonumber\\
&&\!\!\!
- \left(x^2 + (2 - x)^2 \frac{\Delta^2}{4M^2}\right)
{\cal E}_1 {\cal E}_1^\ast
- x^2 \frac{\Delta^2}{4M^2}
\widetilde{{\cal E}}_1 \widetilde{{\cal E}}_1^\ast \Bigg],
\\
\nonumber\\
\label{def-DVCS-LP}
|{\cal T}_{\rm DVCS}|^2_{\rm LP}
&=&
-\frac{2 \lambda \Lambda (2 - y) }{y(2 - x)^2 {\cal Q}^2}
\Bigg[
4(1 - x) \left({\cal H}_1 \widetilde {\cal H}_1^\ast
+ \widetilde{\cal H}_1 {\cal H}_1^\ast \right)
- x^2
\left(
{\cal H}_1 \widetilde{{\cal E}}_1^\ast
+ \widetilde{{\cal E}}_1 {\cal H}_1^\ast +
\widetilde{{\cal H}}_1 {\cal E}_1^\ast
+ {\cal E}_1 \widetilde{{\cal H}}_1^\ast
\right)
\nonumber\\
&&- x \left( \frac{1}{2} x^2 + (2-x) \frac{\Delta^2}{4 M^2} \right)
\left( {\cal E}_1 \widetilde{{\cal E}}_1^\ast
+ \widetilde{{\cal E}}_1 {\cal E}_1^\ast \right)
\Bigg],
\\
\label{def-DVCS-TP}
|{\cal T}_{\rm DVCS}|^2_{\rm TP}
&=&
\frac{8 \cos({\mit\Phi} - \phi_r/2) ( 2 - 2 y + y^2 ) \sqrt{1 - x}}
{\pi y^2 (2 - x)^2 {\cal Q}^2}
\sqrt{\frac{- \Delta^2}{M^2}} \sqrt{1 - \frac{\Delta^2_{\rm min}}{\Delta^2}}
\nonumber\\
&&\!\!\!
\times
\left\{
(2 - x) \left[({\rm Re}{\cal H}_1)({\rm Im} {\cal E}_1)
- ({\rm Im}{\cal H}_1)({\rm Re} {\cal E}_1)\right]
- x \left[({\rm Re}\widetilde{{\cal H}}_1)({\rm Im} \widetilde{{\cal E}}_1)
- ({\rm Im}\widetilde{{\cal H}}_1)({\rm Re} \widetilde{{\cal E}}_1)
\right]
\right\}
\nonumber\\
&&\!\!\!+
\frac{2 \lambda \sin({\mit\Phi} - \phi_r/2) (2 - y) \sqrt{1 - x}}
{\pi y (2 - x)^2 {\cal Q}^2}
\sqrt{- \frac{\Delta^2}{M^2}}
\sqrt{1 - \frac{\Delta^2_{\rm min}}{\Delta^2}}
\Big[ 2 x \left( {\cal H}_1 {\widetilde {\cal E}}_1^\ast
+ {\cal H}_1^\ast {\widetilde {\cal E}}_1 \right)
\\
&&\!\!\! - 2(2 - x) \left( {\widetilde {\cal H}}_1 {\cal E}_1^\ast
+ {\widetilde {\cal H}}_1^\ast {\cal E}_1 \right)
+ x^2 \left( {\cal E}_1 {\widetilde  {\cal E}}_1^\ast
+ {\cal E}_1^\ast {\widetilde {\cal E}}_1 \right) \Big] .
\nonumber
\end{eqnarray}
For the BH term we use an analogous decomposition with
\begin{eqnarray}
\label{cal-BH-LO-unp}
|{\cal T}_{\rm BH}|^2_{\rm unp} &=&
-\frac{2 ( 2 - 2 y + y^2 )}{(1 - y) \Delta^2}
\Bigg[
4 \frac{1 - x}{x^2}
\left(1 -\frac{\Delta_{\rm min}^2}{\Delta^2}\right) F_1^2
+  2 ( F_1 + F_2 )^2
+ \left( \frac{\Delta^2}{\Delta_{\rm min}^2} - 1 \right) F_2^2
\Bigg], \\
\label{cal-BH-LO-LP}
|{\cal T}_{\rm BH}|^2_{\rm LP}
&=& \frac{4\lambda \Lambda (2 - y) y}{(1 - y) \Delta^2}
( F_1 + F_2 ) \Bigg[ 2\frac{1 - x}{x}
\left( 1 - \frac{\Delta_{\rm min}^2}{\Delta^2} \right) F_1
+ F_1 + F_2 \Bigg], \\
\label{cal-BH-LO-TP}
|{\cal T}_{\rm BH}|^2_{\rm TP}
&=& \frac{8 \lambda \sin({\mit\Phi} - \phi_r/2) ( 2 - y ) y
\sqrt{1 - x}}{ \pi (1 - y) x M (-\Delta^2)^{3/2}}
\sqrt{1 - \frac{\Delta^2_{\rm min}}{\Delta^2}} (F_1 + F_2)
\left( 2 x M^2 F_1  +  \Delta^2 F_2 \right) .
\end{eqnarray}
For a polarized lepton beam the interference term is decomposed into a
contribution for an unpolarized and an additional one for a polarized
target, i.e.\
$ {\cal I}^2 \equiv
{\cal T}_{\rm BH} {\cal T}_{\rm DVCS}^\ast +
{\cal T}_{\rm DVCS} {\cal T}_{\rm BH}^\ast
= {\cal I}_{\rm unp}^2(\lambda) + {\cal I}_{\rm pol}^2(\lambda)$,
where
\begin{eqnarray}
\label{def-Intf-unp}
{\cal I}_{\rm unp}^2 (\lambda)
&=& - \frac{8 (2 - 2 y + y^2 ) \sqrt{1 - x}}{
\sqrt{1 - y} y x \sqrt{- \Delta^2 {\cal Q}^2}}
\sqrt{1 -  \frac{\Delta^2_{\rm min}}{\Delta^2}}
\cos(\phi_r) {\rm Re} \Bigg\{F_1 {\cal H}_1
+ \frac{x}{2 - x} ( F_1 + F_2 ) \widetilde{\cal H}_1
\nonumber\\
&&- \frac{\Delta^2}{4 M^2} F_2  {\cal E}_1\Bigg\}
- \frac{8\lambda (2 - y) \sqrt{1 - x}}
{\sqrt{1 - y} x \sqrt{-\Delta^2 {\cal Q}^2}}
\sqrt{1  - \frac{\Delta^2_{\rm min}}{\Delta^2}}
\sin(\phi_r) {\rm Im} \Bigg\{ F_1 {\cal H}_1 \\
&&+ \frac{x}{2 - x} (F_1 + F_2) \widetilde{\cal H}_1
- \frac{\Delta^2}{4 M^2} F_2 {\cal E}_1 \Bigg\},
\nonumber\\
\label{def-Intf-LP}
{\cal I}_{\rm LP}^2(\lambda)
&=& \frac{ 8 \Lambda ( 2 - 2 y + y^2 ) \sqrt{1 - x}}{
\sqrt{1 - y} y x \sqrt{- \Delta^2 {\cal Q}^2}}
\sqrt{1 -  \frac{\Delta^2_{\rm min}}{\Delta^2}}
\sin(\phi_r) {\rm Im}
\Bigg\{
\frac{x}{2 - x} (F_1 + F_2)\left({\cal H}_1 + \frac{x}{2} {\cal E}_1 \right)
\nonumber\\
&&+F_1 \widetilde{\cal H}_1- \frac{x}{2 - x}
\left(
\frac{x}{2} F_1 + \frac{\Delta^2}{4 M^2} F_2 \right)
\widetilde{\cal E}_1 \Bigg\}
+ \frac{8\lambda\Lambda  (2 - y) \sqrt{1 - x}}
{\sqrt{1 - y} x \sqrt{-\Delta^2 {\cal Q}^2}}
\sqrt{1  - \frac{\Delta^2_{\rm min}}{\Delta^2}}
\cos(\phi_r)
\nonumber\\
&&\times
{\rm Re} \Bigg\{ \frac{x}{2 - x} (F_1 + F_2)
\left( {\cal H}_1 + \frac{x}{2} {\cal E}_1 \right)
+F_1 \widetilde{\cal H}_1 - \frac{x}{2 - x}\left(\frac{x}{2}F_1
+\frac{\Delta^2}{4M^2} F_2 \right) \widetilde{\cal E}_1 \Bigg\},
\end{eqnarray}
\begin{eqnarray}
\label{def-Intf-TP}
{\cal I}_{\rm TP}^2(\lambda)
&=&  \frac{8 ( 2 - 2 y + y^2 )}{\sqrt{1 - y} y x (2 - x)
\sqrt{{\cal Q}^2 M^2}}
\Bigg[
\frac{\cos{( {\mit\Phi} - 3 \phi_r/2} )}{2\pi}
\left( 1 -\frac{\Delta^2_{\rm min}}{\Delta^2} \right)
(1 - x)  {\rm Im} \bigg\{
2 F_2 ({\cal H}_1 +{\widetilde{\cal H}}_1)
\nonumber\\
&&
- [ ( 2 - x ) F_1 - x F_2]{\cal E}_1
-x F_1 {\widetilde{\cal E}}_1  \bigg\}
+ \frac{\cos{({\mit\Phi} + \phi_r/2)}}{2\pi}\Bigg(
\frac{\Delta^2_{\rm min}}{\Delta^2}(F_1 + F_2)
{\rm Im} \Big\{4 ( 1 - x )
 \nonumber\\
&&\times \left( {\cal H}_1 - \widetilde {\cal H}_1 \right)
- x^2 ( {\cal E}_1 -\widetilde{\cal E}_1) \Big\}
+ \left( 1 - \frac{\Delta^2_{\rm min}}{\Delta^2} \right)
{\rm Im}\Big\{2 (1 - x) F_2 ({\cal H}_1-\widetilde{{\cal H}}_1)
- [ ( 2 - x ) F_1 \nonumber\\
&&  + x F_2 ]  {\cal E}_1
+x ( F_1 + x F_2 ) \widetilde{\cal{E}}_1 \Big\}\Bigg) \Bigg]
+\frac{ 8\lambda ( 2 - y )}{\sqrt{1 - y} ( 2 - x ) x
\sqrt{ {\cal Q}^2 M^2}}
\Bigg[ \frac{ \sin{ ( {\mit\Phi} -3\phi_r/2 ) }}{2 \pi}
 \\
&&\times \left(1 - \frac{\Delta^2_{\rm min}}{\Delta^2}\right)
( 1 - x ){\rm Re}\Big\{2 F_2 ({\cal H}_1 +{\widetilde{\cal H}}_1)
- [ ( 2 - x ) F_1 - x F_2]{\cal E}_1
-x F_1 {\widetilde{\cal E}}_1 \Big\}
 \nonumber\\
&&- \frac{\sin{({\mit\Phi} + \phi_r/2)}}{2\pi}
  \Bigg(\frac{\Delta^2_{\rm min}}{\Delta^2} ( F_1 + F_2 )
{\rm Re}\Big\{
4 ( 1 - x ) \left({\cal H}_1 -{\widetilde{\cal H}}_1 \right)
- x^2 ({\cal E}_1 - {\widetilde{\cal E}}_1 ) \Big\}
 \nonumber\\
&&+ \left( 1 - \frac{\Delta^2_{\rm min}}{\Delta^2} \right)
{\rm Re} \Big\{ 2 ( 1 - x ) F_2
\left( {\cal H}_1 - {\widetilde{\cal H}}_1 \right)
- [ ( 2 - x ) F_1 + x F_2 ] {\cal E}_1 \nonumber\\
&&+ x ( F_1 + x F_2 ){\widetilde{\cal E}}_1 \Big\}\Bigg) \Bigg] .
\nonumber
\end{eqnarray}

\section{Extraction of leading twist-two amplitudes.}
\label{Sec-ExtofAmp}

A strong motivation for the measurement of DVCS arises from the fact that
the second moment of SPDs in the parity even sector is related to form
factors appearing in the decomposition of the symmetric QCD energy-momentum
tensor ${\mit\Theta}_{\mu\nu}$. A gauge-invariant decomposition of the
matrix element of the angular-momentum tensor, ${\cal M}_{\mu\nu,\sigma}
= x_\mu {\mit\Theta}_{\sigma\nu} - x_\nu {\mit\Theta}_{\sigma\mu}$,
provides, therefore, a separate estimation of the total angular-momentum
fraction carried by quarks and gluons \cite{Ji97}. To achieve this goal
it is necessary to interpolate the corresponding moments of the SPDs to
forward kinematics $\Delta = 0$. The new information that is required (and
not available from DIS) is contained in the SPDs appearing in the
spin-flip amplitude ${\cal E}_1$. Having the spin puzzle in mind, we
give special attention to the problem of extracting ${\cal E}_1$ from
measurements in different kinematical domains.

The simplified explicit expressions (\ref{def-DVCS-unp}-\ref{def-Intf-TP})
for the amplitudes squared allow us to discuss the extraction of the
leading twist-two amplitudes from future experimental data. Moreover,
they allow us to give a qualitative discussion of the ratio for the
different cross sections in more detail. Note, however, that the formulae,
presented below, are only valid in a small kinematical window and in
general not sufficient for numerical estimates. In the following we divide
the kinematical region into $\Delta^2 \approx \Delta_{\rm min}^2$,
$|\Delta_{\rm min}^2| < |\Delta^2| < M^2$, and $ |\Delta_{\rm min}^2|
< M^2 \le |\Delta^2|$. Furthermore, we separately consider the small $x$
region.

\subsection{Small $x$ region.}
\label{SecSubSmax}

Let us start with the small $x$ region, which is for instance relevant for
the HERA experiments \cite{Sau99}. We assume that the small $x$ behavior of
${\cal E}_1 $ and $\widetilde{\cal E}_1 $ is {\it not} one power less than
that of ${\cal H}_1 $ and $\widetilde{\cal H}_1 $, and, furthermore, that
the SPDs in the small $x$ region are essentially determined by the usual
parton densities at least in the DGLAP region. It is easy to establish
that for small values of $x$ the ratio of the real to imaginary part of the
unpolarized amplitude ${\cal H}_1 $ is about 0.3 (by means of dispersion
relation the same magnitude has been obtained in Ref.\ \cite{FraFreStr97}).
In the polarized case we find for instance for the Gehrman-Stirling
parametrization \cite{GehSti96} that this ratio for $\widetilde{\cal H}_1$
is of order $1 - 1.7$. However, the contribution of the latter is quite
small as compared to ${\cal H}_1$.

Taking into account these numbers, we find that for longitudinally polarized
target Eqs.\ (\ref{def-DVCS-unp},\ref{def-DVCS-LP}) can be approximated
by:
\begin{eqnarray}
\label{def-DVCS2-sx}
|{\cal T}_{\rm DVCS}|^2
&\approx& \frac{2(2-2y + y^2)}{y^2 {\cal  Q}^2}
\Bigg[ ({\rm Im}{\cal H}_1)^2- \frac{\Delta^2}{4M^2}
\left\{ ({\rm Im}{\cal E}_1)^2 + ({\rm Re}{\cal E}_1)^2 \right\} \Bigg]
\\&&
- \frac{4\lambda \Lambda (2 - y) }{y {\cal  Q}^2}
\left( {\rm Im} {\cal H}_1 {\rm Im}{\widetilde{\cal H}}_1
+ {\rm Re} {\cal H}_1 {\rm Re}{\widetilde{\cal H}}_1 \right) .
\nonumber
\end{eqnarray}
Contributions containing ${\widetilde {\cal E}}_1$ are down by two powers
of $x$ or they are proportional to $x \frac{\Delta^2}{4M^2}$. The
transversally polarized part offers an opportunity to measure ${\cal E}_1$:
\begin{eqnarray}
|{\cal T}_{\rm DVCS}|^2_{\rm TP}
&\approx& \frac{4 \cos({\mit\Phi} - \phi_r/2) ( 2 - 2 y + y^2 ) }
{\pi y^2 {\cal Q}^2} \sqrt{\frac{- \Delta^2}{M^2}}
\sqrt{1 - \frac{\Delta^2_{\rm min}}{\Delta^2}}
\left[ ({\rm Re}{\cal H}_1)({\rm Im} {\cal E}_1)
- ({\rm Im}{\cal H}_1)({\rm Re} {\cal E}_1) \right]
\nonumber\\
&&\!\!\!-
\frac{4 \lambda \sin({\mit\Phi} - \phi_r/2) (2 - y)}
{\pi y {\cal Q}^2} \sqrt{- \frac{\Delta^2}{M^2}}
\sqrt{1 - \frac{\Delta^2_{\rm min}}{\Delta^2}}
\left[({\rm Im}{\widetilde {\cal H}}_1) ({\rm Im} {\cal E}_1)
+ ({\rm Re}{\widetilde {\cal H}}_1) ({\rm Re} {\cal E}_1) \right].
\nonumber\\
\end{eqnarray}

The interference term starts with $x^{-1}$, however, since the sea quark
and gluonic contributions to the DVCS amplitudes are expected to grow with
$x^{-1}$, the DVCS cross section and the interference term have the same
$x$ dependence. For the longitudinally polarized case it reads for small
$x$:
\begin{eqnarray}
{\cal I}^2 &=& - \frac{8 \sin(\phi_r)
\sqrt{ 1 - \frac{\Delta^2_{\rm min}}{\Delta^2} }}
{\sqrt{1 - y} y x \sqrt{- \Delta^2 {\cal Q}^2}}
{\rm Im}\Bigg\{
 \lambda (2 - y) y \left( F_1 {\cal H}_1 - \frac{\Delta^2}{4 M^2}
F_2 {\cal E}_1\right)
 -\Lambda ( 2 - 2 y + y^2 )
\nonumber\\
&&\times\left(\frac{x}{2}(F_1+F_2){\cal H}_1 +
             F_1 \widetilde{\cal H}_1\right)
\Bigg\}
- \frac{8 \cos(\phi_r) \sqrt{ 1 - \frac{\Delta^2_{\rm min}}{\Delta^2} }}
{\sqrt{1 - y} y x \sqrt{- \Delta^2 {\cal Q}^2}}
{\rm Re} \Bigg\{
( 2 - 2 y + y^2 )
\\
&&\times\left( F_1 {\cal H}_1 - \frac{\Delta^2}{4 M^2}
F_2 {\cal E}_1 \right)
- \lambda\Lambda (2 - y) y  \left(\frac{x}{2}(F_1+F_2){\cal H}_1 +
             F_1 \widetilde{\cal H}_1\right)
 \Bigg\} .
\nonumber
\end{eqnarray}
The interference term might give in future an opportunity to access the
imaginary and real part of the linear combination $F_1 {\cal H}_1 -
\frac{\Delta^2}{4 M^2} F_2 {\cal E}_1 $ due to the charge and single lepton
spin asymmetries for $-\Delta^2_{\rm min} < -\Delta^2$. For a polarized
proton beam one may extract in an analogous way ${\rm Im}
\left(x(F_1+F_2){\cal H}_1/2 + F_1 \widetilde{\cal H}_1\right)$ and
in combination with a polarized lepton beam one also gets the real part of
this expression. Note that one has now the opportunity, at least in
principle, to extract the DVCS cross section (\ref{def-DVCS2-sx}) for
unpolarized or double spin flip experiments and thus separate ${\cal H}_1$
from the ${\cal E}_1$ contributions for the imaginary and real part.

The squared of the BH term will generally start with $x^{- 2}$ (however,
the spin dependent part goes only like $x^{- 1}$) and has therefore a
similar $x$ dependence as the other ones, i.e.\ $x {\rm Im} {\cal H}_1/F_i$
is of order one or so, and grows only slightly with increasing $x$. However,
in comparison with the squared DVCS term it is multiplied by $y^2
\cQ^2/\Delta^2$. Thus, one expects that the BH background is not large in
the small $y$ and $-\Delta^2/\cQ^2$ region. We conclude from our discussion
that the measurement of the unpolarized and longitudinal polarized cross
sections as well as single spin asymmetries it is feasible to disentangle
the imaginary and real part of ${\cal H}_1$, ${\widetilde{\cal H}}_1$, and
${\cal E}_1$ at small $x$ and $-\Delta^2$ of the order of one ${\rm GeV}^2$
or larger. For smaller values ${\cal E}_1$ is in general kinematically
reduced. Since $\widetilde{{\cal E}_1}$ is suppressed at least by one power
in $x$, we conclude that this amplitude will not be accessible in the small
$x$ region.

As mentioned, if $-\Delta^2$ starts to be smaller, the contributions
proportional to ${\cal E}_1$ die out. In the case of its value at the
kinematical boundary, i.e.\ $\Delta^2 \approx \Delta_{\rm min}^2$, we see
that Eqs.\ (\ref{cal-BH-LO-unp}-\ref{cal-BH-LO-TP}) are quite simple, and,
for instance, for a longitudinally polarized target we obtain
\begin{eqnarray}
\label{cal-BH-LO-DeltaMin}
|{\cal T}_{\rm BH}|^2 =
-\frac{4(2-2y + y^2)}{(1 - y) \Delta_{\rm min}^2} (F_1+F_2)^2
+ \frac{4\lambda \Lambda (2 - y) y}{(1 - y) \Delta_{\rm min}^2}
(F_1 + F_2)^2 .
\end{eqnarray}
Moreover, in this case the interference term for unpolarized or
longitudinally polarized target drops out, too. It is worth to have a
closer look at the ratio of the DVCS and BH cross section. The unpolarized
cross section is essentially governed by the imaginary part of
unpolarized parton distributions:
\begin{eqnarray}
\frac{d\sigma_{\rm DVCS}(\lambda = 0, \Lambda = 0)}
{d\sigma_{\rm BH}(\lambda = 0, \Lambda = 0)}
= \frac{(1-y) x^2   M^2}{2  y^2 {\cal Q}^2}
\frac{
{\rm Im}{\cal H}_1^2}{(F_1 + F_2)^2}.
\end{eqnarray}
It is remarkable that the ratio of double spin flip (DF) cross sections
is proportional to the product of unpolarized and polarized parton
distributions:
\begin{eqnarray}
\frac{d\sigma_{\rm DVCS}(\lambda = 1, \Lambda = 1)
- d\sigma_{\rm DVCS}(\lambda = - 1, \Lambda = 1)}
{d\sigma_{\rm BH}(\lambda = 1, \Lambda = 1)
- d\sigma_{\rm BH}(\lambda = - 1,\Lambda = 1)}
= \frac{(1 - y) x^2 M^2}{y^2 {\cal Q}^2}
\frac{{\rm Im}{\cal H}_1  {\rm Im}\widetilde{\cal H}_1
+ {\rm Re}{\cal H}_1  {\rm Re}\widetilde{\cal H}_1 }{(F_1 + F_2)^2}.
\end{eqnarray}

For the sake of completeness we want also to mention that the interference
term for a transversally polarized target does not vanish for $\Delta^2
\approx \Delta_{\rm min}^2$, while both the DVCS and BH cross section are
reduced to the unpolarized ones. This interference term is
proportional to ${\cal H}_1 - {\widetilde{\cal H}}_1$:
\begin{eqnarray}
\label{def-Intf-TP-sm-x}
{\cal I}^2(\lambda)
&=& \frac{16 ( 2 - 2 y + y^2 )}{\sqrt{1 - y} y x \sqrt{{\cal Q}^2 M^2}}
\frac{\cos{({\mit\Phi} + \phi_r/2)}}{2\pi}
(F_1 + F_2)
{\rm Im}\left\{ {\cal H}_1 - \widetilde {\cal H}_1 \right\}
\\
&&- \frac{ 16 \lambda ( 2 - y )} {\sqrt{1 - y} x\sqrt{ {\cal Q}^2 M^2}}
\frac{\sin{({\mit\Phi} + \phi_r/2)}}{2\pi} ( F_1 + F_2 )
{\rm Re}\left\{ {\cal H}_1 -{\widetilde{\cal H}}_1 \right\} .
\nonumber
\end{eqnarray}
Note that at the kinematical boundary the spin flip contributions are
generally suppressed by at least a factor $x^2$. Thus, in this kinematical
range it is hopeless to extract ${\cal E}_1$ or $\widetilde{{\cal E}}_1$
from any experiment.

\subsection{Asymmetries for $|\Delta^2| \gg |\Delta_{\rm min}^2|$.}

Away from the kinematical boundary $|\Delta^2| \gg |\Delta_{\rm min}^2|$
we can consider different cross sections to get separately information
on the real and imaginary parts of the leading twist-two amplitudes,
${\cal H}_1, \dots ,\widetilde {\cal E}_1$. Since the BH cross section
depends only on the product of the lepton and target polarization,
i.e.\ on $\lambda \Lambda$ or $\lambda f({\mit\Phi})$ with $f({\mit\Phi}
+ \pi)= - f({\mit\Phi})$, we have a direct access to the interference
term in polarized experiments to leading order in $1/\sqrt{{\cal Q}^2}$.
Moreover, the relative sign of the interference term is determined by the
charge of the lepton beam. Thus, the following cross sections allow one
to get access to the leading twist-two structure functions in the DVCS
amplitude. In the approximation used above and neglecting terms
${\cal O} \left(\Delta_{\rm min}^2/\Delta^2\right)$, they read:
\begin{enumerate}

\item Polarized positron beam and unpolarized target:
\begin{eqnarray}
\label{def-SSA-SL}
\Delta_{\rm SL} d\sigma
&\equiv& d\sigma^{\uparrow} - d\sigma^{\downarrow}
\\
&=& - \frac{16 (2 - y) \sqrt{1 - x}}
{\sqrt{1 - y} x \sqrt{ - \Delta^2 {\cal Q}^2}}
\sin(\phi_r)
{\rm Im} \Bigg\{ F_1 {\cal H}_1
+ \frac{x}{2 - x} (F_1 + F_2) \widetilde{\cal H}_1
- \frac{\Delta^2}{4 M^2} F_2 {\cal E}_1 \Bigg\} d\cM .
\nonumber
\end{eqnarray}

\item Unpolarized positron beam and longitudinally polarized target:
\begin{eqnarray}
\label{def-SSA-SLN}
\Delta_{\rm SLN} d\sigma &\equiv& d\sigma_{\uparrow}
- d\sigma_{\downarrow}
=  \frac{16 ( 2 - 2 y + y^2 )
\sqrt{1 - x}}{\sqrt{1 - y} y x \sqrt{ - \Delta^2 {\cal Q}^2}}
\sin(\phi_r) \\
&&\!\!\!\times {\rm Im} \left\{
\frac{x}{2 - x} (F_1 + F_2)\left( {\cal H}_1 + \frac{x}{2} {\cal E}_1\right)
+F_1 \widetilde{\cal H}_1
+ \frac{x}{2 - x} \left( \frac{x}{2} F_1
+ \frac{\Delta^2}{4 M^2} F_2 \right) \widetilde{\cal E}_1 \right\}
d \cM .
\nonumber
\end{eqnarray}

\item Unpolarized positron beam and transversally polarized target:
$(\Phi=\{0,\pi\})$:
\begin{eqnarray}
\label{def-SSA-STN}
\Delta_{\rm STN} d\sigma
&\equiv&
d \sigma_{\rightarrow} - d \sigma_{\leftarrow}
=\frac{16(2 - 2 y + y^2)}{\sqrt{1 - y} y x (2 - x)
\sqrt{{\cal Q}^2 M^2}} \nonumber\\
&\times&\Bigg[ \frac{\cos{(3\phi_r/2})}{2\pi}
(1 - x) {\rm Im} \left\{2 F_2 ({\cal H}_1 + {\widetilde{\cal H}}_1)
- [ ( 2 - x ) F_1 - x F_2]{\cal E}_1
-x F_1 {\widetilde{\cal E}}_1\right\}
\nonumber\\
&&+ \frac{\cos{(\phi_r/2)}}{2\pi}
{\rm Im}\Big\{
2 ( 1 - x ) F_2
\left( {\cal H}_1 - {\widetilde{\cal H}}_1 \right)
- [ ( 2 - x ) F_1 + x F_2 ] {\cal E}_1
\\
&& \hspace{9cm}
+ x ( F_1 + x F_2 ){\widetilde{\cal E}}_1
\Big\}
\Bigg] d\cM.
\nonumber
\end{eqnarray}

\item Charge asymmetry in unpolarized experiment:
\begin{eqnarray}
\label{def-CAsy-unp}
\Delta_{\rm C}^{\rm unp} d\sigma
&\equiv& d {^{+}\!\sigma}^{\rm unp} - d{^{-}\!\sigma}^{\rm unp}
= - \frac{16 ( 2 - 2 y + y^2 ) \sqrt{1 - x}}{
\sqrt{1 - y} y x \sqrt{- \Delta^2 {\cal Q}^2}}
\cos(\phi_r)
\\
&&\times
{\rm Re} \left\{F_1 {\cal H}_1
+ \frac{x}{2 - x} (F_1+F_2) \widetilde{\cal H}_1
- \frac{\Delta^2}{4 M^2} F_2 {\cal E}_1\right\} d\cM . \nonumber
\end{eqnarray}

\item Charge asymmetry in double spin-flip experiments with longitudinally
polarized target:
\begin{eqnarray}
\label{def-CAsy-LP}
\Delta_{\rm C}^{\rm DFL} d\sigma &\equiv&
d {^{+}\!\sigma}_{\uparrow}^{\uparrow}
- d{^{-}\!\sigma}_{\downarrow}^{\downarrow}
- \Delta_{\rm C}d\sigma^{\rm unp}
= \frac{16 (2 - y) \sqrt{1 - x}}
{\sqrt{1 - y} x \sqrt{-\Delta^2 {\cal Q}^2}}
\cos(\phi_r) \\
&&\times{\rm Re} \Bigg\{ \frac{x}{2 - x} (F_1 + F_2)
\left( {\cal H}_1 + \frac{x}{2} {\cal E}_1 \right)
+F_1 \widetilde{\cal H}_1 -\frac{x}{2 - x} \left( \frac{x}{2}F_1
+  \frac{\Delta^2}{4M^2} F_2 \right)
\widetilde{\cal E}_1 \Bigg\} d\cM . \nonumber
\end{eqnarray}

\item Charge asymmetry in double spin-flip experiments with transversally
polarized target:
\begin{eqnarray}
\label{def-CAsy-TP}
\Delta_{\rm C}^{\rm DFT} d\sigma &\equiv &
d {^{+}\!\sigma}_{\rightarrow}^{\uparrow}
- d{^{-}\!\sigma}_{\leftarrow}^{\downarrow} -
\Delta_{\rm C} d\sigma^{\rm unp}
=
- \frac{ 8( 2 - y )}{\sqrt{1 - y} ( 2 - x ) x \sqrt{ {\cal Q}^2 M^2}}
\nonumber\\
&\times& \Bigg[ \frac{ \sin{ (3\phi_r/2 ) }}{2 \pi}
2 ( 1 - x ){\rm Re}\Big\{2 F_2 ({\cal H}_1 + {\widetilde{\cal H}}_1)
- [ ( 2 - x ) F_1 - x F_2]{\cal E}_1
-x F_1 {\widetilde{\cal E}}_1 \Big\}
\nonumber\\
&&+ \frac{\sin{(\phi_r/2)}}{\pi}
{\rm Re} \Big\{ 2 ( 1 - x ) F_2
\left( {\cal H}_1 - {\widetilde{\cal H}}_1 \right)
- [ ( 2 - x ) F_1 + x F_2 ] {\cal E}_1
\\
&& \hspace{9cm}
+x ( F_1 + x F_2 ){\widetilde{\cal E}}_1 \Big\}
\Bigg]
d\cM , \nonumber
\end{eqnarray}
\end{enumerate}
where $d\cM = \frac{\alpha^3 x y} { 8 \, \pi \, {\cal Q}^2}
\left( 1 + \frac{4 M^2 x^2}{{\cal Q}^2} \right)^{-1/2} dx d y
d|\Delta^2| d\phi_r$. Note that for a consequent numerical treatment the
$\Delta_{\rm min}^2/\Delta^2$ dependence cannot be droped and can be
easily restored from Eqs.\ (\ref{def-Intf-unp}--\ref{def-Intf-TP}).

As we see from this list, the two single spin cross sections of
longitudinally polarized beam or target give us information on the imaginary
part of two linear combinations of the four leading twist-two amplitudes.
The real part of these two quantities is accessible via charge asymmetry in
unpolarized or longitudinally double spin flip experiments. Obviously, at
low $|\Delta^2|$, i.e.\ compared to $4 M^2 \approx 4\ {\rm GeV}^2$, there is
a suppression of contributions proportional to $E$ and $\tilde E$ SPDs,
which are theoretically not well constraint: the contribution proportional
to $E$ SPDs completely drops, while $\widetilde E$ is suppressed by $x^2$.
For the kinematics of present experiments $- \Delta^2 \sim 1 {\rm GeV}^2$,
and thus it is not a good approximation to drop afore mentioned terms,
$\Delta^2/4 M^2 \sim \cO (1)$. However, in certain asymmetries these
contributions are accompanied by a factor $x$ and thus can be safely
discarded at smaller values of $x$. Further constraints separately for
the imaginary and real part arise from spin and charge asymmetries for
a transversally polarized target. Fortunately, due to different angular
dependences, we have indeed two further independent linear combinations
of the DVCS amplitudes. But the kinematical pre-factors and an additional
angular dependence of the denominator for large $y$, which we have
dropped for simplicity, makes their practical use questionable.

Finally, the whole cross section for unpolarized and polarized $\gamma$
production tests the leading twist-two approximation used above. However,
to be consistent one has then to expand the BH and interference terms up to
order $1/\cQ^2$. As outlined in section \ref{Sec-CrosSec} this is
straightforward for the BH cross section, while in the interference term new
twist-three contributions will enter. These additional terms will be worked
out in a forthcoming paper. Here we should only mention that the different
azimuthal angle dependence can be used to pick up different combinations of
helicity amplitudes, entering at different twist levels. For instance, for
small $y$ it is justified to use the derived formulae which imply that
all single spin and charge asymmetry cross sections integrated over the
azimuthal angle $\phi_r$ vanish at twist-two level, for instance,
\begin{equation}
\int_0^{2\pi} d\phi_r
\frac{\Delta_{\rm SL} d\sigma}{d\phi_r} =
{\cal O} \left(1/{\cal Q}^2\right)
\quad
\mbox{\ or\ }
\quad
\int_0^{2\pi} d\phi_r
\frac{\Delta_{\rm C}^{\rm unp} d\sigma}{d \phi_r} =
{\cal O} \left(1/{\cal Q}^2\right).
\end{equation}

\section{Numerical estimates.}
\label{Sec-NumEst}

To give phenomenological predictions for the asymmetries discussed
above we have to specify models for the SPDs. Definitely, this is the
main source of uncertainty for the numerical estimates we present in
this section. (Of course, the primary goal of experiments is rather to
constrain the skewed functions via the theoretical formulae.)

Let us discuss in turn spin non-flip and flip functions. In the former
case SPDs have a definite limit for forward kinematics when they reduce
to the familiar parton densities. For spin non-flip SPDs we choose
an oversimplified factorized form of the $\Delta^2$ and $(t,\xi)$
dependence of the SPDs:
\begin{equation}
\label{AnsSPDH}
H^i (t, \xi, \Delta^2, \cQ^2) = F^i_1 (\Delta^2) q^i (t, \xi, \cQ^2) ,
\qquad
{\widetilde H}^i (t, \xi, \Delta^2, \cQ^2)
=  G^i_1 (\Delta^2) \Delta q^i (t, \xi, \cQ^2) ,
\end{equation}
where $i$ denotes the quark flavour. Here $F^i_1 (\Delta^2)$ and
$G^i (\Delta^2)$ are elastic parton form factors normalized to unity
at the origin, $F_1^i (0) = G^i (0) = 1$, and $q (t, \xi,\cQ^2)$ as
well as $\Delta q (t, \xi,\cQ^2)$ are the non-forward functions
specified below. The support of these functions is $[-1,1]$, where
for $t > 0$ we have the quark distribution and for $t < 0$ the antiquark
distribution, i.e.\ $\bar q (t, \xi, {\cal Q}^2) = - q (- t, \xi,
{\cal Q}^2)$ and $\Delta \bar q (t, \xi, {\cal Q}^2) = \Delta q (-t,
\xi, {\cal Q}^2)$. The normalization of $H^i (t, \xi, \Delta^2,
{\cal Q}^2)$ and ${\widetilde H}^i (t, \xi, \Delta^2, {\cal Q}^2)$
at $\Delta^2 = 0$, $\xi = 0$ is determined by parton densities $H^i
(t, 0, 0) = q^i (t)$, $\widetilde{H}^i (t, 0, 0) = \Delta q^i (t)$.
Since the $\xi$ dependence drops out in the first moment we can constrain
the $\Delta^2$ dependence by sum rules, e.g.\
\begin{equation}
\label{Def-SumRul}
\int_{-1}^1 dt\, H^i (t, \xi, \Delta^2) = F^i_1 (\Delta^2) ,
\qquad
\int_{-1}^1 dt\, E^i (t, \xi, \Delta^2) = F^i_2 (\Delta^2) ,
\end{equation}
with the Dirac and Pauli form factor, respectively. For non-polarized SPDs
the valence $u$ and $d$ quark form factors in the proton can be easily
deduced from (\ref{dipoleFF}) via $F_I^{({p \atop n})} = 2 \left( {Q_u
\atop Q_d} \right) F_I^u + \left( {Q_d \atop Q_u} \right) F_I^d$ which
results in
\begin{equation}
2 F^u_I (\Delta^2) = 2 F_I^p (\Delta^2) + F_I^n (\Delta^2),
\qquad
F^d_I (\Delta^2) = F_I^p (\Delta^2) + 2 F_I^n (\Delta^2) ,
\quad\mbox{for}\quad I = 1,2.
\end{equation}
For $s$ (or in general, sea) quark contribution in the parity even sector
the first moment vanishes and the sum rule (\ref{Def-SumRul}) does not give
any constraint. Nevertheless, the counting rule for elastic form factors
tells us that for large $\Delta^2$ we have a $(\Delta^2)^{-3}$ behaviour.
This suggests the following dipole fit with the mass cutoff $m_V$ chosen
as for valence quarks:
\begin{equation}
\label{SQFF}
G_E^{\rm sea} (\Delta^2)
= \frac{1}{1 + \kappa_{\rm sea}} G_M^{\rm sea} (\Delta^2)
= \left( 1 - \frac{\Delta^2}{m_V^2} \right)^{- 3},
\end{equation}
where $\kappa_{\rm sea}$ will be specified below.

So far the model has been governed by the known forward densities.
Unfortunately, a similar reduction is absent for the helicity flip
amplitudes. For $E^i$ we adopt nevertheless
\begin{equation}
E^i (t, \xi, \Delta^2) = r^i (t, \xi) F^i_2 (\Delta^2)
\quad
\mbox{with}
\quad
r^i (t, \xi) = q^i (t, \xi).
\end{equation}
The identification of $r^i (t, \xi)$ with $q^i (t, \xi)$ ensures the sum
rule for valence quark contributions. Note that the parameter $\kappa_{\rm
sea}$ normalizes the sea quark contribution, for instance, $\kappa_{\rm sea}
= 0$ provides $E^{\rm sea} = 0$. In the axial vector channel the quark form
factors can be read off from the iso-triplet axial form factor $G_1^{(3)}
(\Delta^2)$ of the $\beta$-decay and related by isotopic symmetry to the
form factor in question. The decay constant $g_A^{(3)}$ is expressed by
Goldberger-Treiman relation $g_A^{(3)} \approx \frac{1}{\sqrt{2}M} f_\pi
g_{\pi NN}$ in terms of the pion decay constant $f_\pi$, nucleon mass $M$
and $\pi NN$-coupling, $g_{\pi NN}$, and has the numerical value $g_A^{(3)}
= 1.267$ \cite{Oku90}. If we assume the same $\Delta^2$-dependence for the
iso-singlet $G_1^{(0)} (\Delta^2)$ with the same cutoff and a constant
$g_A^{(0)}$ we get for quarks
\begin{equation}
G^i_1 (\Delta^2) = \left( 1 - \frac{\Delta^2}{m_A^2} \right)^{- 2},
\quad\mbox{for}\quad i = \{ u_v, d_v \},
\quad\mbox{and}\quad
G^{\rm sea}_1 (\Delta^2) = \left( 1 - \frac{\Delta^2}{m_A^2} \right)^{- 3}
\end{equation}
with the scale $m_A = 0.9\, {\rm GeV}$ \cite{Oku90}.

Finally for the polarized spin-flip amplitude it was observed
\cite{Rad98,RadManPil98,FraPolStr99} that, similar to the $\beta$-decay
effective pseudoscalar form factors \cite{Oku90}, one can approximate
$\widetilde E$ at small $\Delta^2$ by the pion pole so that one ends up
with the model
\begin{equation}
\widetilde E^u = \widetilde E^d
= \frac{1}{2\xi} \theta \left( \xi - |t| \right)
\phi_\pi \left( t/\xi \right) g_\pi (\Delta^2),
\quad\mbox{with}\quad
g_\pi (\Delta^2) = \frac{4 g_A^{(3)} M^2}{m_\pi^2 - \Delta^2}
\quad\mbox{and}\quad
\phi_\pi (x) = \frac{4}{3} (1 - x^2) .
\end{equation}

It is reliable to assume that the SPDs in the DGLAP region can be modelled
by the forward parton distributions measured in inclusive reactions. In
the simplest case we assume that this is the only contribution. We refer
to this model as the forward parton distribution (FPD) model which has no
skewedeness dependence at the input scale $\cQ_0$:
\begin{equation}
(\Delta) q^i (t, \xi, {\cal Q}^2_0) = (\Delta) q^i (t, {\cal Q}^2_0)
\quad\mbox{for}\quad t > 0.
\end{equation}
The contributions for $t < 0$ are easily restored by means of symmetry.
Although the input distribution does not depend on $\xi$ a small evolution
step does generate such a $\xi$-dependence \cite{BelMul99}. A further model
is based on an proposal for the so-called double distribution (DD)
 \cite{MueRobGeyDitHor94}, namely,
\begin{equation}
\label{nontodouble}
q (t, \xi, Q^2) = \int_{-1}^1 dx \int_{- 1 + |x|}^{1 - |x|} dy
\delta ( x + \xi y - t ) f ( y, x, Q^2 ) .
\end{equation}
The functional dependence in the $x$-subspace is given by the shape of the
forward parton density, $f (x)$, while the $y/[1 - |x|]$-dependence of
the integrand has to be similar to that of the distribution amplitude.
Thus, $f(y, x)$ is given \cite{Rad98} by the product of a forward
distribution $f (z)$ (more precisely $q (z)$ for quarks and $z g (z)$
for gluons) with a profile function $\pi$
\begin{equation}
\label{DD-Ansatz}
f ( y, x ) = \pi ( y, x ) f ( x ),
\end{equation}
where $\pi$ for quarks and gluons is given by
\begin{equation}
\label{profile}
\pi^Q (y, x)
= \frac{3}{4} \frac{ [1 - |x|]^2 - y^2 }{ [1 - |x|]^3 },
\qquad
\pi^G (y, x)
= \frac{15}{16}
\frac{ \left\{ [1 - |x|]^2 - y^2 \right\}^2}{[1 - |x|]^5} .
\end{equation}
Now we are in a position to give numerical estimates for the cross
sections defined in the preceding sections.

\subsection{HERA kinematics.}

In the small $x$ kinematics, $10^{-5} \le x \le 10^{-2}$, first estimates
for the unpolarized azimuthal and single spin asymmetry has been given in
the factorization \cite{FraFreStr97} and BFKL \cite{BalKuc00} approach. In
the following we evaluate numerically different asymmetries in order to
demonstrate that ${\cal H}_1$, $\widetilde{{\cal H}}_1$ and ${\cal E}_1$ are
measurable in HERA experiments. In the following we deal with the FPD model,
where we equate the SPDs with the parton densities taking the MRS A$'$
\cite{MarRobSti95} and the GS A \cite{GehSti96} parametrization at the input
scale $\cQ^2 = 4 \mbox{\ GeV}^2$ and using $\bar{u} = \bar{d} = \bar{s}/2$ as
well as $\Delta\bar{u} = \Delta\bar{d} = \Delta\bar{s}$. For simplicity we
do not discuss the $\cQ^2$ dependence of our predictions due to the
perturbative evolution.

\begin{figure}[t]
\vspace{-0.5cm}
\begin{center}
\mbox{
\begin{picture}(0,140)(300,0)
\put(60,0){\insertfig{8}{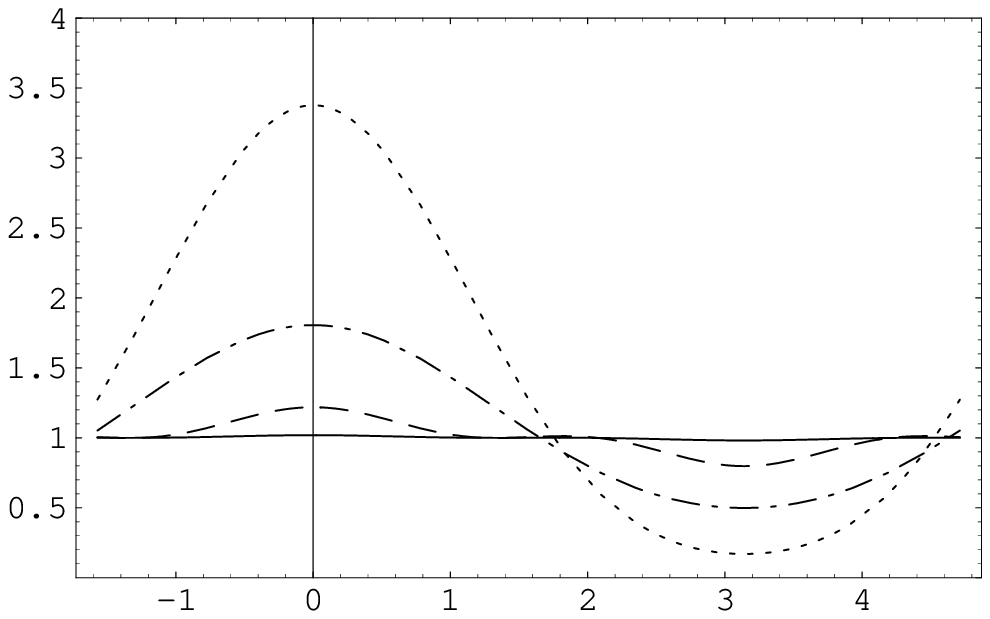}}
\put(200,100){(a)}
\put(280,-10){$\phi_r$}
\put(300,0){\insertfig{8.5}{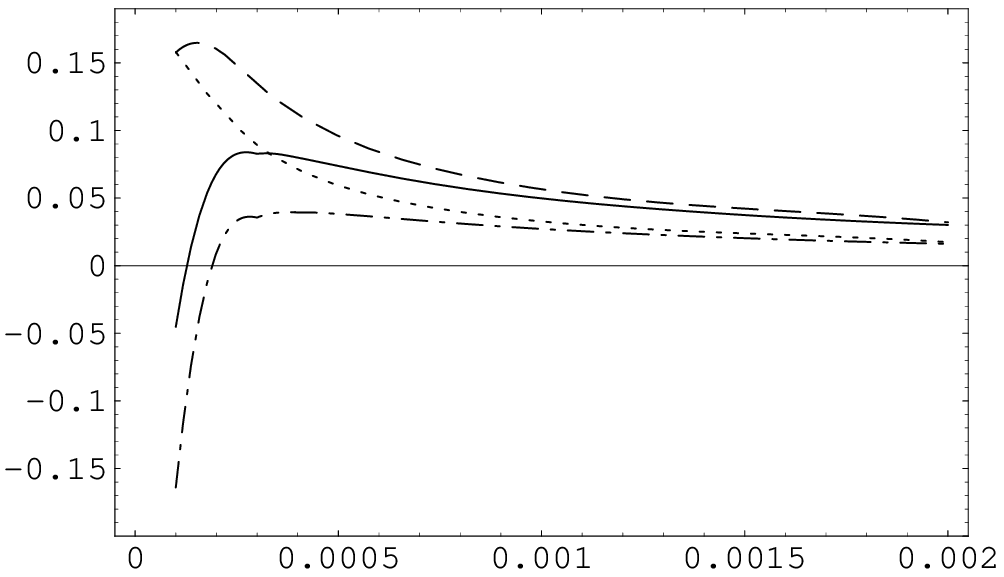}}
\put(400,30){(b)}
\put(530,-10){$x$}
\end{picture}
}
\end{center}
\caption{\label{BHphiDep} (a) Ratios of different approximations of the
BH cross section to the exact one in dependence of the azimuthal angle
$\phi_r$ for $\cQ^2 = 4 \mbox{\ GeV}^2$ and $x = 10^{-4}$ for
the Pade-type approximation for $\Delta^2 = - 0.1 \mbox{\ GeV}^2$ (solid)
and $\Delta^2=-0.5 \mbox{\ GeV}^2$ (dashed) and for the leading
approximation (\ref{cal-BH-LO-unp}) (dash-dotted and dotted, respectively).
(b) Unpolarized azimuthal angle asymmetry for $\Delta^2 = - 0.05 (- 0.25)
\mbox{\ GeV}^2$ versus $x$. Solid (dash-dotted) and dashed (dotted) lines
show the result for the Pade and leading approximation, respectively,
where $\kappa_{\rm sea}$ is set to zero.}
\end{figure}

The unpolarized azimuthal asymmetry is defined by
\begin{eqnarray}
\label{def-AziAsy}
A =
\frac{ \int_{-\pi/2}^{\pi/2} d\phi_r \frac{ d\sigma^{\rm unp}}{d\phi_r }
- \int_{\pi/2}^{3\pi/2} d\phi_r  \frac{d\sigma^{\rm unp}}{d\phi_r } }{
\int_{0}^{2\pi} d \phi_r  \frac{ d\sigma^{\rm unp}}{d\phi_r }} ,
\end{eqnarray}
and its first signature has been seen by the ZEUS collaboration \cite{Sau99}.
In the approximation (\ref{cal-BH-LO-unp}) the unpolarized squared BH term
is $\phi_r$ independent. However, it turned out that this approximation
may provide misleading results due to the neglected azimuthal dependence of
the BH process. In Fig.\ \ref{BHphiDep}(a) it can be seen that the dropped
terms cause a strong $\phi_r$-dependence and that only for the Pade-type
approximation introduced in section \ref{Sec-CrosSec} [see Eq.\
(\ref{app-prop}) for the expansion of the propagator] there is a good
agreement with the exact expression. Since the BH cross section is in
general suppressed in the upper hemisphere, one may expect quite different
predictions depending on the approximations involved. Indeed, for small
values of $x$, where $y = \cQ^2/x s $ with the center of mass energy $s
= 4 \cdot 27.5 \cdot 820 \mbox{\ GeV}^2$, we find a strong deviation,
even for small $-\Delta^2$. Note that the interference term is only
taken into account up to order $1/\sqrt{-\Delta^2 \cQ^2}$ and at present
it is not clear whether the $1/\cQ^2$ term is crucial or not.

\begin{figure}[ht]
\vspace{-1cm}
\begin{center}
\mbox{
\begin{picture}(0,300)(300,0)
\put(60,150){\insertfig{8}{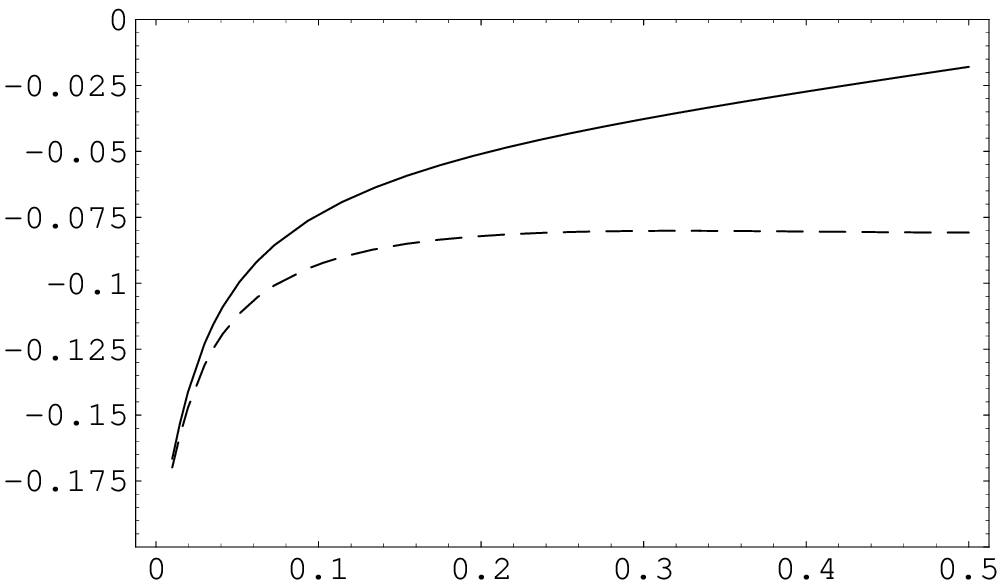}}
\put(130,180){(a)}
\put(370,180){(b)}
\put(300,150){\insertfig{8}{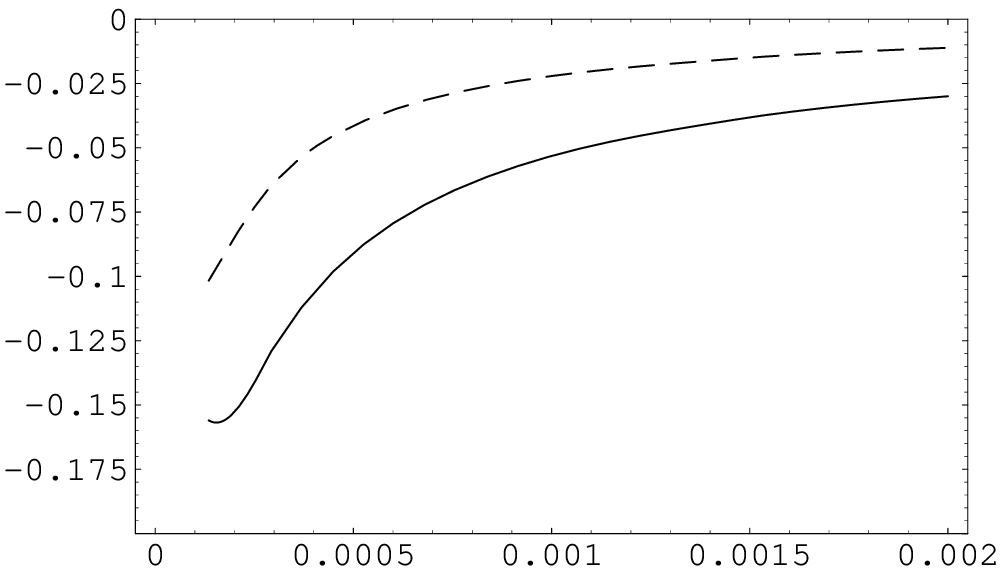}}
\put(60,0){\insertfig{8}{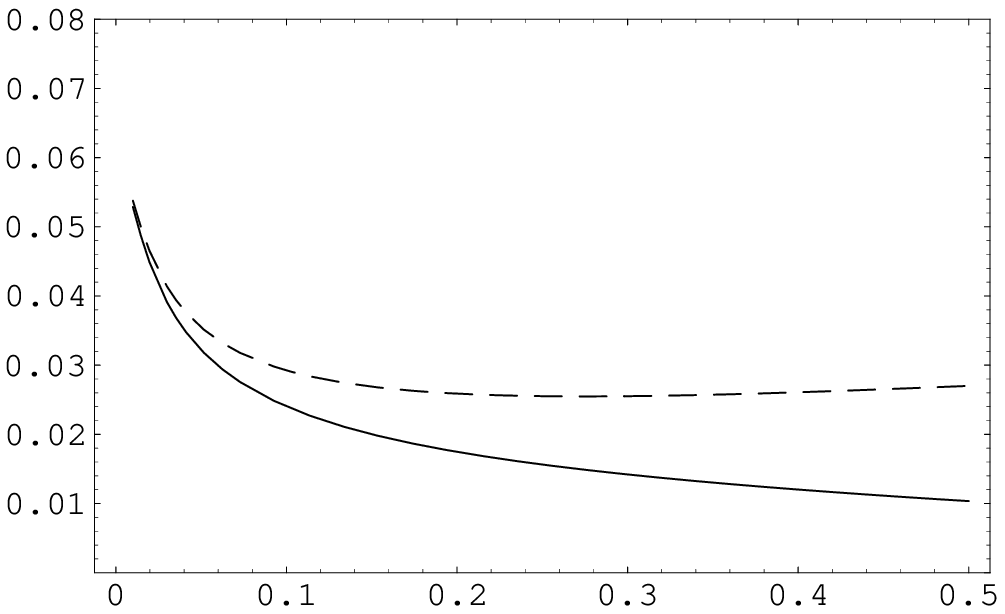}}
\put(300,0){\insertfig{8}{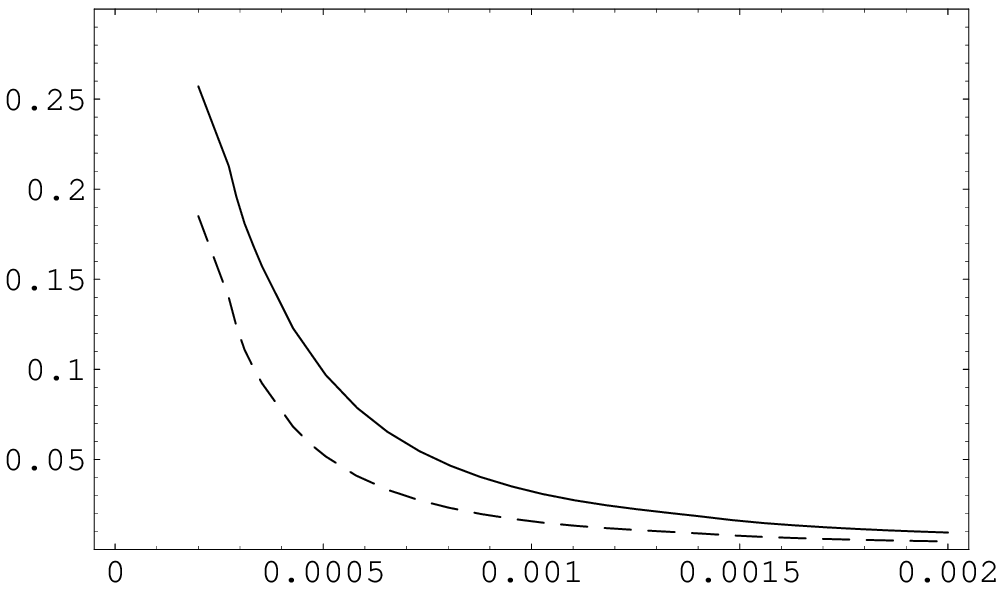}}
\put(130,110){(c)}
\put(370,110){(d)}
\put(280,-10){$-\Delta^2$}
\put(530,-10){$x$}
\end{picture}
}
\end{center}
\caption{\label{ChaSpiAziAsy} Unpolarized azimuthal charge asymmetry
$A_{\rm C}$ for $\cQ^2 = 4 \mbox{GeV}^2$ and $x = 5\cdot 10^{-4}$ with
$\kappa_{\rm sea} = - 2$ at LO for the complete expression (solid curve)
and neglecting the ${\cal E}_1$ contribution (dashed curve) plotted in
(a) versus $- \Delta^2$. The same asymmetry for $\Delta^2 = - 0.05
\mbox{\ GeV}^2$ (solid curve) and $\Delta^2 = - 0.25 \mbox{\ GeV}^2$
(dashed curve) at LO is shown in (b) for the region $1\cdot 10^{-4}
\le x \le 2\cdot 10^{-3}$. In (c) and (d) the same is shown for the electron
single spin asymmetry.}
\end{figure}

To get information on $\left\{F_1 {\cal H}_1 - \frac{\Delta^2}{4 M^2} F_2
{\cal E}_1\right\}$ in a cleaner way one may use the azimuthal asymmetry
of the charge asymmetry $\Delta_C^{\rm unp} d\sigma =
d{^{-}\!\sigma}^{\rm unp}-d{^{+}\!\sigma}^{\rm unp}$
defined in Eq.\ (\ref{def-CAsy-unp}) (in the following we restore the
$\Delta^2_{\rm min}/\Delta^2$ dependence in the interference amplitudes):
\begin{eqnarray}
\label{ChargeAs}
A_{\rm C}
= \frac{ \int_{-\pi/2}^{\pi/2} d\phi_r
\frac{\Delta_C^{\rm unp}  d\sigma}{d\phi_r}
- \int_{\pi/2}^{3\pi/2} d\phi_r \frac{\Delta_C^{\rm unp} d\sigma}{d\phi_r} }{
\int_{0}^{2\pi} d\phi_r
\frac{d{^{-}\!\sigma}^{\rm unp} +  d{^{+}\!\sigma}^{\rm unp}}{d\phi_r}} ,
\end{eqnarray}
and the single (lepton) spin asymmetry with unpolarized target
$\Delta_{\rm SL} = d{\sigma}^{\uparrow} -  d{\sigma}^{\downarrow}$
defined in Eq.\ (\ref{def-SSA-SL}):
\begin{eqnarray}
\label{SingleAs}
A_{\rm SL}
= \frac{ \int_{0}^{\pi} d\phi_r \frac{\Delta_{\rm SL} d\sigma}{d\phi_r}
- \int_{\pi}^{2\pi} d\phi_r  \frac{\Delta_{\rm SL} d\sigma}{d\phi_r} }{
\int_{0}^{2\pi} d\phi_r
\frac{d{\sigma}^{\uparrow} + d{\sigma}^{\downarrow}}{d\phi_r}} .
\end{eqnarray}
The former (later) one is proportional to the real (imaginary) part of
the linear combination that was mentioned. Both of them give sizeable
effects. Although $A_{\rm SL}$ is proportional to the imaginary part which
is growing as $x^{-1}$ this asymmetry is suppressed by a kinematical factor
$y$ as compared to the charge asymmetry. Therefore, we find in the
considered kinematics a two times larger value of the charge asymmetry
as for the single spin one. This should be reversed for larger values of
$y$ (see also the discussion in Ref.\ \cite{FreStr99}). In Fig.\
\ref{ChaSpiAziAsy} we demonstrate the influence of the amplitude
${\cal E}_1$ which is quite sizeable for larger values of $-\Delta^2$,
where $-\Delta^2/\cQ^2$ remains small.

To measure $\widetilde{\cal H}_1$ one have to consider the proton single
spin asymmetry $\Delta_{\rm SLN} = d{\sigma}_{\uparrow} -
d{\sigma}_{\downarrow}$ given in Eq.\ (\ref{def-SSA-SLN}):
\begin{eqnarray}
A_{\rm SLN}
= \frac{ \int_{0}^{\pi} d\phi_r \frac{\Delta_{\rm SLN} d\sigma}{d\phi_r}
- \int_{\pi}^{2\pi} d\phi_r \frac{\Delta_{\rm SLN} d\sigma}{d\phi_r} }{
\int_{0}^{2\pi} d\phi_r
\frac{d{\sigma}_{\uparrow} + d{\sigma}_{\downarrow}}{d\phi_r}}
\end{eqnarray}
and the charge asymmetry in double spin-flip experiments, i.e.\
$\Delta_{\rm C}^{\rm DFL} d\sigma = d {^{-}\!\sigma}_{\uparrow}^{\uparrow}
- d{^{+}\!\sigma}_{\downarrow}^{\downarrow} - \Delta_{\rm C}d\sigma^{\rm
unp}$, defined in Eq.\ (\ref{def-CAsy-LP})
\begin{eqnarray}
A_{\rm C}^{\rm DFL}
= \frac{ \int_{0}^{\pi} d\phi_r
\frac{\Delta_{\rm C}^{\rm DFL} d\sigma}{d\phi_r}
- \int_{\pi}^{2\pi} d\phi_r
\frac{\Delta_{\rm C}^{\rm DFL} d\sigma}{d\phi_r} }{
\int_{0}^{2\pi} d\phi_r
\frac{d {^{-}\!\sigma}^{\rm unp} + d{^{+}\!\sigma}^{\rm unp}}{d\phi_r}} .
\end{eqnarray}
Unfortunately, for the considered $\cQ^2$ value we find for our model that
this single spin asymmetry is compatible with zero. Although, we have an
enhancement for small $y$ due do the factor $1/y$ in comparison to the
lepton single spin asymmetry, this enhancement cannot compensate the
weaker rise of $\widetilde{{\cal H}}_1$ for small $x$. For larger value
of $\cQ$, i.e.\ also larger $x$, and $-\Delta^2$ and small $y$ we found a
few percent effect. Note also that the ratio of proton to lepton single
spin asymmetry gives us the ratio $({\rm Im}\widetilde{{\cal H}}_1)/
( {\rm Im}{\cal H}_1)$ times $1/y$, which is sizeable at not too small
$x$. The azimuthal asymmetry in double spin flip experiments normalized to
the unpolarized cross section is tiny. However, if we would compare it with
the extracted double spin flip part, we would get quite sizeable effects
which are of the order of 10\% or even more. Furthermore, to explore
${\cal H}_1$ separately, one can make use of the discussion given in
section \ref{SecSubSmax}. Especially, for small $y$ the BH cross section
is suppressed by $y^2$ in comparison to the DVCS one and, thus, the former
one drops out.

Let us note that the perturbative NLO corrections to the imaginary part for
small $x$ are of order of 20\% for each quark species as well as for the
gluon contribution.

\subsection{HERMES kinematics.}

Now let us turn to the HERMES experiment with a $E = 27.5\ {\rm GeV}$
positron beam scattered by a hydrogen target and give predictions for
the asymmetries which can be accesed there. To give numerical
predictions we take the model for SPDs deduced from the double
distribution model and forward (un-) polarized parton densities from
Ref.\ (\cite{GRV95}) \cite{GRSV97} with $\kappa_{\rm sea}=-2.$ As a
starting point we choose ${\cal Q}^2 = 4\ {\rm GeV}^2$, $x$-range
$0.1 - 0.4$, and the $t$-channel momentum transfer $- \Delta^2 =
0.1 - 0.5\ {\rm GeV}^2$. Since the parton densities are defined at
rather low momentum scale ${\cal Q}_0^2 = 0.23\ {\rm GeV}^2$ we evolve
them up to the experimental scale using the formalism of \cite{BelMul98}.
We concentrate mostly on the cross sections proportional to the
imaginary part of the DVCS amplitude. Since $\widetilde E_1$ is
concentrated in the ER-BL region with zero at $|t| = \xi$, ${\rm Im}
\widetilde {\cal E}_1$ vanishes at all order.

\begin{figure}[t]
\vspace{-1cm}
\begin{center}
\mbox{
\begin{picture}(0,300)(300,0)
\put(60,150){\insertfig{8}{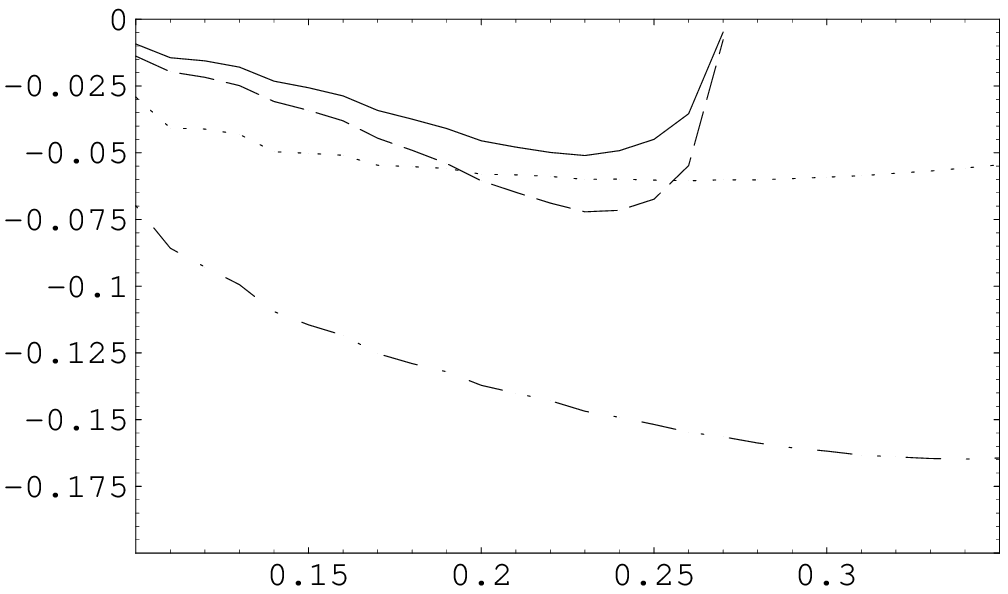}}
\put(100,200){(a)}
\put(350,260){(b)}
\put(300,150){\insertfig{8}{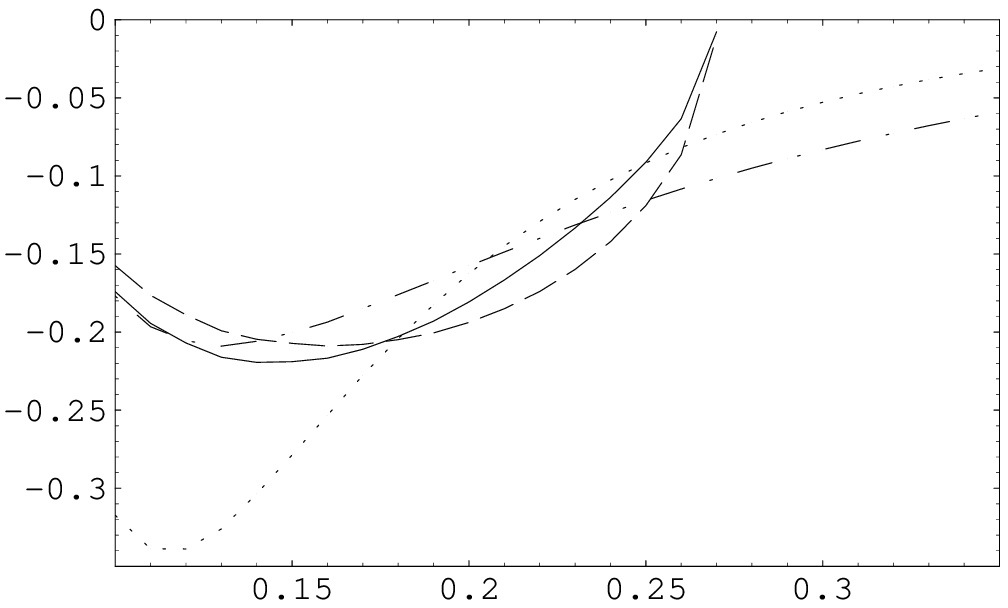}}
\put(60,0){\insertfig{8}{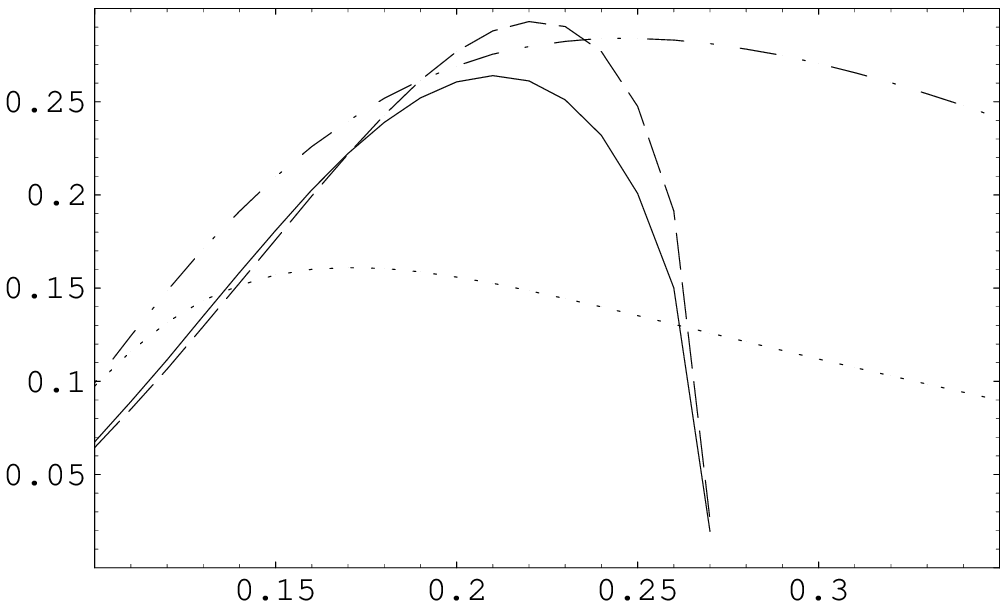}}
\put(300,0){\insertfig{8}{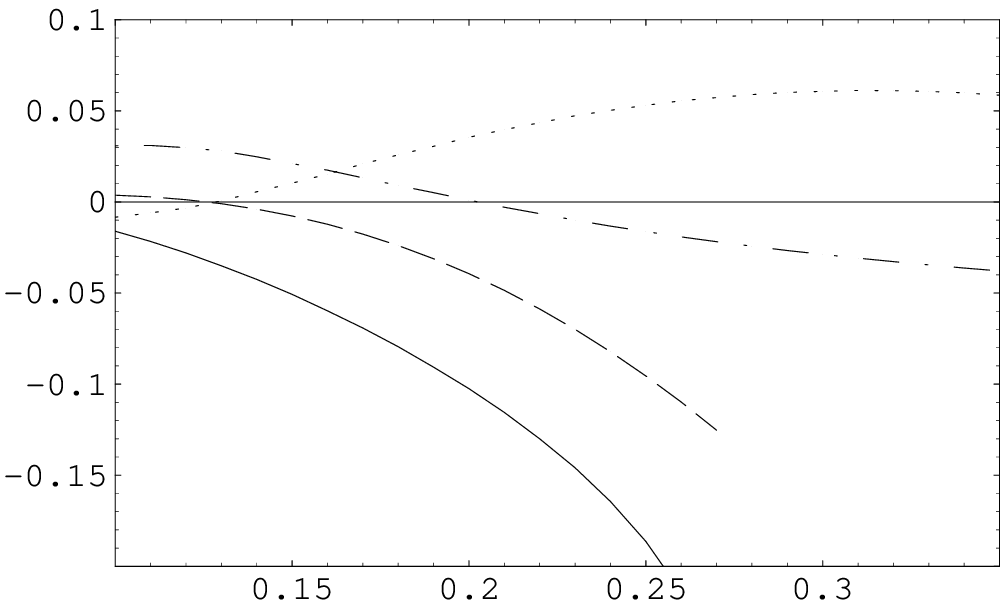}}
\put(100,100){(c)}
\put(350,40){(d)}
\put(280,-10){$x$}
\put(530,-10){$x$}
\end{picture}
}
\end{center}
\caption{\label{Fig-HERMES} Perturbative leading order results for the
charge asymmetry for an unpolarized beam (a), single spin asymmetries
for a polarized positron beam (b) and an unpolarized target; as well as
for an unpolarized lepton beam and a longitudinally (c) (transversally (d))
polarized proton target versus $x$, for $\cQ^2 = 4\ {\rm GeV}^2$. The
predictions for the model specified in the text are shown as solid (dotted)
curves for $\Delta^2 = -0.1 (0.5)\ {\rm GeV}^2$, respectively. The same
model however with neglected spin-flip contributions are presented as
dashed (dash-dotted) line for the same values of $\Delta^2$.}
\end{figure}

Similarly to the previous section we calculate the charge asymmetry for
unpolarized experiments given in Eq.\ (\ref{ChargeAs}). As we see form the
Fig.\ \ref{Fig-HERMES}(a) it reaches the level of $5-15\%$ for $- \Delta^2 =
0.5\ {\rm GeV}^2$. The single lepton spin asymmetry (\ref{SingleAs}) is much
larger and can be as big as $20-30\%$ [see Fig.\ \ref{Fig-HERMES}(b)] which
gives promises to measure it experimentally. Note that in both cases the
cross sections are very sensitive to the helicity flip distribution $E$
which is responsible for the orbital momentum of partons in the proton.

In view of limited space let us consider finally only the spin asymmetry
for a longitudinally and transversally polarized proton (\ref{def-SSA-STN}).
In the former case azimuthal averaging of Eq.\ (\ref{def-SSA-SLN}) is the
same as in Eq.\ (\ref{SingleAs}) and leads to a sizeable asymmetry of order
20\%. In the later case in order to extract the combination of SPDs
multiplied by $\cos{(\phi_r/2)}$ we define the following azimuthal asymmetry
\begin{eqnarray}
A_{\rm STN}
= \frac{
\int_{-\pi/3}^{2\pi/3} d\phi_r
\frac{\Delta_{\rm STN} d\sigma}{d\phi_r}
- \int_{2\pi/3}^{5\pi/3} d\phi_r
\frac{\Delta_{\rm STN} d\sigma}{d\phi_r} }{
\int_{0}^{2\pi} d\phi_r
\frac{
d \sigma_{\rightarrow} + d \sigma_{\leftarrow}}{d\phi_r} }.
\end{eqnarray}
The numerical estimate presented in Fig.\ \ref{Fig-HERMES} (d) demonstrates
that it has a sizable effect which in contrast to the other symmetries does
not vanish at the kinematical boundery $\Delta^2=\Delta^2_{\rm min}$.

\section{Conclusions.}
\label{Sec-Con}

In this paper we have given theoretical predictions for diverse asymmetries
which can be measured in exclusive leptoproduction experiments of a real
photon. Our estimates are rather encouraging, since they demonstrate a
possibility to separate the contributions coming from different leading
twist-two DVCS amplitudes by means of polarized lepton beam, and
longitudinally and transversally polarized targets. This is the first step
on the way to constrain the form of the SPDs from experimental data. The
models used for numerical estimates lead to large charge and single
(lepton) spin asymmetries of order $20\%$. However, since an asymmetry
gives information on a linear combination of SPD one has to resort to
other combinations of cross sections in order to disentangle a given
distribution.

We have not discussed in the present paper the NLO correction to the DVCS
amplitudes. However, as has been shown in Ref.\ \cite{BelMulNieSch00} the
latter could be very sizable and therefore change the LO predictions
significantly in the valence quark region. Yet another important issue
is the study of kinematical and dynamical higher twist corrections. Both
of these problems deserve a detailed investigation and will be considered
elsewhere.

\vspace{0.5cm}

We would like to thank M.~Amarian, M.~Diehl, and A.V.~Radyushkin for
constuctive correspondence. This work was supported by DFG and BMBF. D.M.
thanks G. Sterman for the hospitality extended to him at the C.N. Yang
Institute for Theoretical Physics while this work was finished.

\end{document}